**Estimating Complier Average Causal Effects for Clustered RCTs
When the Treatment Affects the Service Population**

Peter Z. Schochet, Ph.D.
Senior Fellow
Mathematica
P.O. Box 2393
Princeton NJ, 08542-2393
pschochet@mathematica-mpr.com

May 2022

**Abstract**

RCTs sometimes test interventions that aim to improve existing services targeted to a subset of individuals identified after randomization. Accordingly, the treatment could affect the composition of service recipients and the offered services. With such bias, intention-to-treat estimates using data on service recipients and nonrecipients may be difficult to interpret. This article develops causal estimands and inverse probability weighting (IPW) estimators for complier populations in these settings, using a generalized estimating equation approach that adjusts the standard errors for estimation error in the IPW weights. While our focus is on more general clustered RCTs, the methods also apply (reduce) to non-clustered RCTs. Simulations show that the estimators achieve nominal confidence interval coverage under the assumed identification conditions. An empirical application demonstrates the methods using data from a large-scale RCT testing the effects of early childhood services on children's cognitive development scores.

# 1. Introduction

In clustered randomized controlled trials (RCTs), groups (such as schools, hospitals, or communities) are randomized to treatment and control conditions rather than individuals. Clustered RCTs are often used to test interventions that are impractical to be randomly assigned to individuals and are becoming increasingly prevalent in social policy research [1-2] and medical trials [3].

Clustered RCTs sometimes test interventions that aim to improve existing services targeted to a subset of individuals in the study clusters with special needs or interests. Because service receipt is typically identified after randomization, the size and composition of the service receipt samples may be affected by the intervention and therefore differ in the treatment and control clusters. For example, consider an education RCT with school-level randomization that tests the effects of an intervention that provides enhanced training to teacher aides who tutor third-grade students identified in need of assistance during the school year [4]. Because the enhanced training is intended to improve tutoring practices, it could also affect the size and characteristics of the tutored students and the nature of the tutoring services offered in the treatment clusters. We refer to these types of treatment effects as *service receipt effects*.

In RCTs with service receipt effects, it may be possible to collect data on a pre-specified intention-to-treat (ITT) population at risk of receiving services, which includes persons who ultimately receive services and those who do not. For instance, in our tutoring example, the ITT population could be pre-specified as all third-grade students in the study schools or those with low achievement scores in the prior year. With data on an ITT sample, an RCT can meet industry evidence standards—such as from the What Works Clearinghouse (WWC) for education RCTs [5] or from the Consolidated Standards of Reporting Trials (CONSORT) statement for medical trial [6]—if the study can establish baseline equivalence of individuals and clusters in the analytic treatment and control group samples and meet attrition requirements.

Yet, even with a balanced ITT sample, standard ITT estimators that compare mean outcomes of the full treatment and control groups can be difficult to interpret in the presence of service receipt effects. In these cases, ITT estimators could confound (i) "pure" intervention effects for the average service recipient, (ii) compositional effects due to differences in service recipients across the treatment and control clusters, and (iii) potential spillover effects from service recipients to nonrecipients within the same cluster.

Further, the resulting ITT estimators may have low power for detecting intervention effects if service take-up is low. Such dilution could occur for several reasons. First, the study might lack prior information and data to limit the ITT population to those likely to receive services. Second, RCT implementation considerations may require a broad ITT sample, for example, to facilitate study buy-in from cluster-level staff and study subjects.

Accordingly, with service receipt effects, the interpretation of the ITT findings can be enhanced by estimating treatment effects for well-defined complier populations of service recipients. This article develops estimands and estimators in these settings. We assume no other forms of ITT sample distortions, often referred to in the medical RCT literature as "recruitment biases" [7-15]. These biases could include entry effects into the study clusters or treatment-control differences in study consent rates, which could arise because treatment assignments are rarely blinded to cluster staff and individuals. Thus, we consider RCT settings with balanced ITT samples, but unbalanced service receipt samples.

Our context may be realistic, for example, for smaller-scale interventions that comprise only a part of the total service experience (so are unlikely to affect cluster choices and study consent), and for outcomes measured in the shorter term (before knowledge of the intervention and its effects become more widespread). Our setting, however, may be less realistic for RCTs of larger-scale interventions that involve major changes to cluster-level operations and services.

Relatedly, we focus on interventions that enhance existing offered services rather than those that introduce new service structures, so that individuals in the local service areas at risk of entering the study clusters are not differentially referred to the treatment clusters (which could occur, for instance, if hospitals in the treatment group offer a new medical treatment). This condition also ensures that the relevant populations of targeted service recipients can be identified in both the treatment and control clusters.

We develop complier average causal effect (CACE) estimands by separating the ITT population into principal strata using a potential outcomes framework, where the principal strata are defined by individuals' would-be service receipt statuses in the treatment and control conditions. We develop inverse probability weighting (IPW) estimators [16-18] that pertain to various policy-relevant complier populations and that rely on ignorability conditions for consistency. Thus, our approach provides an alternative to (i) principal stratification methods that rely on distributional assumptions on potential outcomes [19] and (ii) instrumental variable (IV) methods that rely on monotonicity and exclusion restrictions [20] that may not apply in our setting. We also discuss specification tests for the logit propensity score models used to generate the IPW weights. We use generalized estimating equation (GEE) estimation methods [21] that incorporate estimation error in the weights by extending the methods in [22-24] to our clustered RCT setting. We conduct simulations to assess confidence interval coverage and conduct a case study using data from a large-scale RCT assessing the effects of early childhood services on children's cognitive development [25]. R code for estimation is provided in the Appendix.

While our analysis is motivated by service receipt bias for clustered RCTs, our methods also apply (reduce) to non-clustered RCTs with service receipt bias where both research groups can receive related services. Our empirical case study demonstrates such broader applications.

The rest of the article is in seven sections. Section 2 discusses the related literature that this work builds on. Section 3 presents notation and the potential outcomes framework. Section 4 presents our focal causal estimands and Section 5 presents the IPW estimators. Section 6 presents simulation results and Section 7 presents a case study. Section 8 concludes.

## 2. Related Literature

Our analysis builds on the previous literature on methods to adjust for sources of treatment-control group imbalance in clustered RCTs. Li et al. [26] use a principal stratification (PS) approach to define causal estimands for clustered RCTs when the recruited RCT samples differ for the treatment and control groups. However, this study does not develop estimators or consider estimands for RCTs with service receipt effects only (our focus). Donner and Klar [9] reference the PS estimator developed in Frangakis and Rubin [19] to address recruitment bias in general, but do not provide details on estimands or estimation methods. Schochet [27] develops a PS estimator for a well-defined estimand when recruitment bias is due to mobility effects in longitudinal clustered RCTs, but not due to service receipt effects. Cheng and Small [28] and Schochet [29] use a PS approach to discuss well-defined CACE estimators for RCTs with multiple treatment arms, but not with service receipt bias.

Leyrat et al. [30] conduct simulations to examine the extent to which propensity score methods (including IPW estimators) can reduce recruitment bias in clustered RCTs. They generate simulated RCT data with recruitment bias that yields unbalanced ITT samples, and then estimate treatment effects using propensity score methods that balance the treatment and control samples using the simulated baseline covariates. Their study results, however, are difficult to interpret as it is unclear what causal ITT or CACE estimands are being estimated [25]. In contrast, we develop formal estimators for well-defined CACE estimands for RCTs in the special case where recruitment bias is due to service receipt effects only.

Our methods are also motivated by the RCT literature on the use of propensity score methods to estimate intervention effects for subgroups defined by treatment group members' post-randomization intervention experiences [31-32]. These methods apply to settings where service subgroups (e.g., those who receive a particular service package from a program) are defined for the treatment group but are missing for the control group who do not receive comparable services (or whose service receipt is considered part of the service counterfactual). In contrast, in our setting, service receipt is defined for both research groups rather than for the treatment group only. Relatedly, our work is motivated by the RCT literature on propensity score weighting methods to estimate mediating mechanisms through which a treatment exerts an impact on an outcome in the presence of treatment-mediator interactions [33-34]. The focus of these methods, however, is to produce consistent estimators of direct and indirect mediator effects, rather than to estimate treatment effects for service compliers in the presence of service receipt bias.

### 3. Notation and Potential Outcomes Framework

Consider a clustered RCT of *m* total clusters, where $m^1 = mp$ are randomly assigned to the treatment group and $m^0 = m(1-p)$ are randomly assigned to the control group ($0 < p < 1$). Let $T_j = 1$ for clusters in the treatment group and $T_j = 0$ for clusters in the control group.

We assume a well-defined ITT sample—with selection criteria pre-specified in study protocols (e.g., RCT registries)—that contains the pool from which the service recipients will be identified after randomization. The ITT sample can consist of all individuals in the study clusters or a pre-specified subset, but it must contain (1) all persons (or a random sample) who ultimately receive services (to obtain causal estimands for well-defined populations) and (2) some of their peers who do not receive services (to allow for propensity score modeling of the service receipt decision). To make it explicit that our estimands are conditional on those in the ITT population only, we define the $ITT_{ij}$ indicator that equals 1 if individual $i$ in cluster $j$ is in the ITT sample and 0 otherwise. We assume the ITT sample contains $n_j$ persons in cluster $j$, with $n^1$ and $n^0$ persons in the full treatment and control groups, and $n = n^1 + n^0$ persons overall.

We assume baseline, outcome, and service receipt data are available for the ITT sample (e.g., from school or hospital records or study administered assessments and surveys). Further, we assume service receipt bias is evident based on differences across research groups in service receipt rates and the baseline characteristics of service recipients (see [35] for a discussion of imbalance tests). We also assume baseline equivalency of the ITT samples has been established.

We conceptualize the service receipt decision as binary. Let $R_{ij}(T_j)$ denote the potential service receipt status for a person in the $ITT_{ij} = 1$ population, where $R_{ij}(1) = 1$ if the person would receive services in the treatment condition and 0 otherwise, and similarly for $R_{ij}(0)$ in the control condition. Importantly, we assume $R_{ij}(T_j)$ is defined over a class of services related to the intervention (e.g., tutoring services offered in both the treatment and control schools), where services could be received in the study clusters as well as elsewhere. We do not separate $R_{ij}(T_j)$ into multiple service receipt indicators, because while this approach could yield more nuanced estimands, it could lead to large numbers of principal strata with challenges in identifying complier average causal effects. For similar reasons, we assume low rates of controls who receive direct treatment services to avoid principal strata devoted to such crossovers.

We use similar notation to define the potential outcomes. Let $Y_{ij}(1) = Y_{ij}(1, R_{ij}(1))$ be a person's potential outcome if assigned to a treated cluster and let $Y_{ij}(0) = Y_{ij}(0, R_{ij}(0))$ be the potential outcome in the control condition. Potential outcomes are assumed to be continuous.

Our analysis relies on three assumptions. The first is the stable unit treatment value assumption (SUTVA) [36] that we adapt to our clustered RCT context.

*Assumption 1 (A1): SUTVA*: $R_{ij}(T_j)$ and $Y_{ij}(T_j)$ depend only on the person's cluster treatment assignment and not on the assignments of other clusters in the sample. Further, $R_{ij}(T_j)$ and $Y_{ij}(T_j)$ do not depend on the service receipt (compliance) of other persons within or across clusters. SUTVA also assumes a particular unit cannot receive different forms of the treatment.

SUTVA assumes no peer effects. In clustered RCTs, the condition of no between-cluster spillover is likely to be plausible in settings where there are few interactions between individuals across clusters. Within-cluster spillover may be more common. However, the standard potential outcomes and IV framework breaks down in this setting [37] and the identification and estimation of causal effects becomes complex [38]. Note that the no-spillover condition is not unique to our analysis: it pertains to any clustered RCT with noncompliance. With spillover, the potential outcomes framework becomes approximate.

Under SUTVA, the data generating processes for the observed outcome measure, $y_{ij}$, and the observed service receipt status, $r_{ij}$, are a consequence of the assignment mechanism:

$$y_{ij} = T_j Y_{ij}(1) + (1 - T_j) Y_{ij}(0) \tag{1a}$$

$$r_{ij} = T_j R_{ij}(1) + (1 - T_j) R_{ij}(0). \tag{1b}$$

The relation in (1a) states that we can observe $y_{ij} = Y_{ij}(1)$ for those in the treatment group and $y_{ij} = Y_{ij}(0)$ for those in the control group, but not both, and similarly for $r_{ij}$ in (1b).

Our second condition is the complete randomization of clusters [39].

*(A2): Complete randomization of clusters*: Each cluster has the same assignment probability to the treatment group, so that $T_j$ is independent of $R_{ij}(1)$, $R_{ij}(0)$, $Y_{ij}(1)$, and $Y_{ij}(0)$.

Our third condition assumes balanced ITT samples across the treatment and control groups.

*(A3): No recruitment bias in the ITT samples*: There are no treatment effects on the composition of study participants in the ITT samples, for instance, due to entry effects into the study clusters or treatment-control differences in study consent rates.

Additional assumptions for obtaining consistent IPW estimators are presented in Section 5 after we discuss our focal estimands in Section 4.

## 4. Estimands

The ITT estimand, $\tau_{ITT}$, is the average treatment effect in the ITT population:

$$\tau_{ITT} = E[Y_{ij}(1) - Y_{ij}(0) | ITT_{ij} = 1]. \tag{2}$$

Here, it is assumed that individual-level treatment effects, $(Y_{ij}(1) - Y_{ij}(0))$, are weighted equally in the ITT sample, so that cluster-level treatment effects, $(\bar{Y}_j(1) - \bar{Y}_j(0))$ are weighted by their ITT sample sizes, where $\bar{Y}_j(t) = \frac{1}{n_j} \sum_{i=1}^{n_j} Y_{ij}(t)$ are mean potential outcomes for $t \in \{1,0\}$. Other weighting options exist, such as equal cluster weighting, which can be incorporated into the analysis by adjusting the weights [40].

To develop causal estimands for service compliers, we define principal strata based on the four possible $(R_{ij}(1), R_{ij}(0))$ service combinations for those in the ITT population (Figure 1). The (1,1) stratum includes service recipients in both research conditions (always-takers), the (0,0) stratum includes never-takers, and the (1,0) and (0,1) strata include those who would receive services in one research condition but not the other. Without further conditions, we cannot determine the specific principal stratum for any individual, because service receipt is observed in one research condition only (so we only observe the margins in Figure 1).

**Figure 1. Principal stratification framework for service receipt decisions**

|  | Service receipt in the control condition ($R_{ij}(0)$) | | Population share |
|---|---|---|---|
| Service receipt in the treatment condition ($R_{ij}(1)$) | 1 | 0 |  |
| 1 | $\mu^1_{11}, \mu^0_{11}, \tau_{11}, \pi_{11}$ | $\mu^1_{10}, \mu^0_{10}, \tau_{10}, \pi_{10}$ | $\pi_{11} + \pi_{10}$ |
| 0 | $\mu^1_{01}, \mu^0_{01}, \tau_{01}, \pi_{01}$ | $\mu^1_{00}, \mu^0_{00}, \tau_{00}, \pi_{00}$ | $\pi_{01} + \pi_{00}$ |
| Population share | $\pi_{11} + \pi_{01}$ | $\pi_{10} + \pi_{00}$ | 1 |

Notes. The parameters, $\mu^1_{rr'}$ and $\mu^0_{rr'}$ are mean population outcomes in the treatment and control conditions in principal stratum $(r, r')$ for $r, r' \in \{1,0\}$; $\tau_{rr'} = \mu^1_{rr'} - \mu^0_{rr'}$ are population average treatment effects; and $\pi_{rr'}$ are stratum shares that sum to 1.

The average treatment effect for those in principal stratum $(r, r')$ can be defined as

$$\tau_{rr'} = \mu^1_{rr'} - \mu^0_{rr'} = E[Y_{ij}(1) - Y_{ij}(0)|R_{ij}(1) = r, R_{ij}(0) = r', ITT_{ij} = 1] \quad (3)$$

for $r, r' \in \{1,0\}$. Thus, the ITT estimand in (2) is a weighted average of the four $\tau_{rr'}$ estimands:

$$\tau_{ITT} = \pi_{11}\tau_{11} + \pi_{10}\tau_{10} + \pi_{01}\tau_{01} + \pi_{00}\tau_{00}, \quad (4)$$

where $\pi_{rr'}$ are stratum shares that sum to 1. SUTVA implies that $\tau_{00} = 0$.

For our analysis, we focus on three estimands of potential policy interest that pertain to different complier populations and service contrasts. The first estimand pertains to average treatment effects for compliers who would receive services in the treatment (T) condition regardless of their service receipt in the control (C) condition—which we refer to as the CACE-T estimand. Using (3) and (4), this estimand can be obtained by calculating a weighted average of the $\tau_{rr'}$ parameters across the (1,1) and (1,0) populations:

$$\tau_{CACE-T} = \frac{1}{\pi_{CACE-T}}(\pi_{11}\tau_{11} + \pi_{10}\tau_{10}), \quad (5)$$

where $\pi_{CACE-T} = \pi_{11} + \pi_{10} > 0$. This estimand captures intervention effects relative to status quo services for compliers likely to be served if the intervention was rolled out more broadly. It requires weaker identifying conditions for estimation than our other estimands (see Section 5).

Our second estimand pertains to average treatment effects for compliers who would receive services in *either* research condition—which we refer to as the CACE-TC estimand. It pertains to average treatment effects for compliers in the combined (1,1), (1,0) and (0,1) populations:

$$\tau_{CACE-TC} = \frac{1}{\pi_{CACE-TC}} (\pi_{11}\tau_{11} + \pi_{10}\tau_{10} + \pi_{01}\tau_{01}), \tag{6}$$

where $\pi_{CACE-TC} = \pi_{11} + \pi_{10} + \pi_{01} > 0$. This estimand captures intervention effects relative to status quo services for the broad population of compliers typically targeted for services and includes effects for those who gain and lose services.

Our final estimand, $\tau_{11}$, pertains to compliers in the (1,1) population who would receive services in *both* the treatment and control conditions. This estimand pertains to pure intervention effects that arise solely due to improved services for the always-takers.

## 5. IPW Estimators

The principal stratification methods discussed in Frangakis and Rubin [19] and Zhang et al. [41] can be used to estimate each $\mu^1_{rr'}$, $\mu^0_{rr'}$, and $\pi_{rr'}$ parameter in (4), which can then be used to estimate our three focal causal estimands. Parameter identification for these methods is driven by distributional assumptions about potential outcomes within each principal stratum, and parameters are typically estimated using maximum likelihood methods for finite mixture models. However, Feller et al. [42] show these methods can yield unstable results if there is weak separation of the $\mu^1_{rr'}$ and $\mu^0_{rr'}$ means across principal strata, which they stress could occur in RCTs where treatment effects are often small, and recommend these methods be supplemented using nonparametric bounds. Further, unstable results are likely to be more severe for clustered RCTs where the degrees of freedom are based on the number of clusters, not subjects.

Instead, we adopt commonly used IPW estimators for our focal estimands that rely on ignorability conditions [16-18]. We require several definitions. First, let $\boldsymbol{X}_{ij}$ be a row-vector of pretreatment covariates (confounders) that can include both cluster- and individual-level variables, where $\boldsymbol{\mathcal{X}}$ denotes the support of the distribution of $\boldsymbol{X}_{ij}$ in the study population. Second, let $e^1(\boldsymbol{X}_{ij})$ and $e^0(\boldsymbol{X}_{ij})$ denote propensity scores in the treatment and control conditions [43], which, in our context, are service receipt probabilities given the covariates:

$$e^t(\boldsymbol{X}_{ij}) = Pr(R_{ij}(t) = 1 | T_j = t, \boldsymbol{X}_{ij}) = Pr(R_{ij}(t) = 1 | \boldsymbol{X}_{ij}) \qquad (7)$$

for $t \in \{1,0\}$ and $\boldsymbol{X}_{ij} \in \boldsymbol{\mathcal{X}}$. Here, we have removed the conditioning on $ITT_{ij} = 1$ to simplify notation, a convention we follow hereafter. The second equality in (7) holds due to (A2).

The key identification assumption needed for our IPW estimators is that $\boldsymbol{X}_{ij}$ contains all confounders related to both service receipt decisions and potential outcomes (so there are no additional, unmeasured confounders). We discuss two such ignorability assumptions—the first one for the $\tau_{CACE\_T}$ estimand and the second, stronger one for the $\tau_{CACE\_TC}$ and $\tau_{11}$ estimands:

*(A4a): Ignorability of service receipt:*

$$Y_{ij}(t') \perp\!\!\!\perp R_{ij}(s) | T_j = t, \boldsymbol{X}_{ij} \qquad (8)$$

for $(t, t', s) \in \{1,0\}$ and $\boldsymbol{X}_{ij} \in \boldsymbol{\mathcal{X}}$.

This assumption states that given the observed pretreatment confounders, potential service receipt is statistically independent of potential outcomes. This means that in each research group, service receipt is random for individuals with the same covariate values. It is possible to relax (A4a) to only require mean independence [44]. Further, for the $\tau_{CACE\_T}$ estimand, we only require (8) to hold for the treatment group, but we use the more general notation.

For the $\tau_{CACE\_TC}$ and $\tau_{11}$ estimands, we require a stronger ignorability condition that assumes the potential service receipt in one research condition provides no information on the potential service receipt in the other research condition, conditional on $T_j$ and the covariates:

***(A4b): Ignorability of service receipt with no interactions*:**

$$Y_{ij}(t') \perp\!\!\!\perp R_{ij}(s)|R_{ij}(1-s) = s', T_j = t, X_{ij} \tag{9}$$

for $(t, t', s, s') \in \{1,0\}$ and $X_{ij} \in \mathcal{X}$.

This means, for instance, that the same propensity score model for $R_{ij}(1)$ applies to those with $R_{ij}(0) = 1$ and $R_{ij}(0) = 0$. Clearly, (A4b) implies (A4a).

We also need a positivity assumption for the propensity scores:

***(A5): Positivity*:** $0 < e^1(X_{ij}), e^0(X_{ij}) < 1$ for $X_{ij} \in \mathcal{X}$.

This common support condition rules out the perfect predictability of service receipt given the covariates [44-45]. It ensures that for all possible values of the covariates, there is a positive probability of receiving services. Note that randomization ensures that $Pr(T_j = t | X_{ij}) > 0$.

Finally, we require a model specification assumption for the propensity scores:

***(A6): Correct model specifications for $e^1(X_{ij})$ and $e^0(X_{ij})$*.**

Note that the true model specification can differ for the treatment and control groups, including the specific covariates that enter the models and the underlying parameter values.

In what follows, we first develop IPW estimators for the $\tau_{CACE-T}$ estimand and then for the $\tau_{CACE-TC}$ and $\tau_{11}$ estimands. For reference, Table 1 summarizes key features of the estimators.

### 5.1. Estimators for the $\tau_{CACE-T}$ Estimand

The differences-in-means IPW estimator, $\hat{\tau}_{CACE-T}$, compares the mean outcomes of treatments who received services (that is, those with $r_{ij} = 1$) to the weighted mean outcomes of the full control group. The control group weights are $w_{ij} = \hat{e}^1(X_{ij})$, where $\hat{e}^1(X_{ij})$ is a consistent estimator of $e^1(X_{ij})$, the propensity score (service receipt probability) in the treatment condition as defined in (7). More formally, $\hat{\tau}_{CACE-T}$, can be expressed as

$$\hat{\tau}_{CACE-T} = \bar{\bar{y}}_W^1 - \bar{\bar{y}}_W^0 = \frac{1}{w^1}\sum_{j=1}^{m}\sum_{i=1}^{n_j} r_{ij}T_j y_{ij} - \frac{1}{w^0}\sum_{j=1}^{m}\sum_{i=1}^{n_j} \hat{e}^1(X_{ij})(1-T_j)y_{ij}, \qquad (10)$$

where $y_{ij}$ is the observed outcome, $w^1 = \sum_{j=1}^{m}\sum_{i=1}^{n_j} r_{ij}T_j$ is the number of treatments with an $r_{ij} = 1$ value and $w^0 = \sum_{j=1}^{m}\sum_{i=1}^{n_j} \hat{e}^1(X_{ij})(1-T_j)$ is the sum of the weights for the controls. More succinctly, we can express the weights as $w_{ij} = T_j r_{ij} + (1-T_j)\hat{e}^1(X_{ij})$.

To provide intuition for this estimator, note first that service recipients in the treatment group are observed to be in the CACE-T population (i.e., in the (1,1) or (1,0) group), so their weights are set to $w_{ij} = 1$. Similarly, the weights for treatment group nonrecipients are set to $w_{ij} = 0$. For controls, however, we can only calculate their probability of membership in the CACE-T population. Thus, we set their weights to $w_{ij} = \hat{e}^1(X_{ij})$, noting that $\hat{e}^1(X_{ij})$ applies to both research groups due to randomization. As discussed in Appendix A (Result 1), under (A1)-(A3), (A4a), (A5)-(A6), and regularity conditions, $\hat{\tau}_{CACE-T}$ is a consistent estimator as $m \to \infty$.

In practice, weighted least squares (WLS) can be used for estimation by regressing $y_{ij}$ on $T_j$ using the $w_{ij}$ weights. Precision can be improved by including in the model a $1xk$ vector of baseline covariates, $X_{ij}$, where we assume a linear specification, $E[y_{ij}|T_j, X_{ij}] = \beta_0 + T_j\tau_{CACE-T} + X_{ij}\beta$, where $\beta_0$, $\tau_{CACE-T}$, and $\beta = (\beta_1, ..., \beta_k)'$ are model parameters. We can also specify separate treatment and control group models [23]. However, we do not adopt this approach because it can seriously reduce the degrees of freedom (and efficiency) for clustered RCTs that often include relatively small numbers of clusters for cost reasons [40,46].

To show that the resulting WLS estimator, $\hat{\tau}_{CACE-T}^{WLS}$, is consistent, note first that

$$\hat{\tau}_{CACE-T}^{WLS} = (\bar{y}_W^1 - \bar{y}_W^0) - (\bar{\bar{X}}_W^1 - \bar{\bar{X}}_W^0)\hat{\beta}, \tag{11}$$

where $\bar{\bar{X}}_{Wv}^1 = \frac{1}{w^1}\sum_{j=1}^{m}\sum_{i=1}^{n_j} r_{ij}T_j X_{ijv}$ is the treatment group mean for covariate $v = 1, \ldots, k$, $\bar{\bar{X}}_{Wv}^0 = \frac{1}{w^0}\sum_{j=1}^{m}\sum_{i=1}^{n_j} \hat{e}^1(X_{ij})(1-T_j)X_{ijv}$ is the weighted control group mean, and $\hat{\beta}$ are parameter estimates [40,47]. If the IPW weights are correctly specified, then $(\bar{\bar{X}}_W^1 - \bar{\bar{X}}_W^0) \xrightarrow{p} 0$, where the symbol $\xrightarrow{p}$ denotes convergence in probability, implying that $\hat{\tau}_{CACE-T}^{WLS} \xrightarrow{p} \tau_{CACE-T}$.[1]

In sum, the WLS estimator, $\hat{\tau}_{CACE-T}^{WLS}$, can be estimated in three steps:

1. Estimate the propensity score model for the treatment group (see Section 5.3).

2. Calculate the weights by setting: (i) $w_{ij} = 1$ for treatments with $r_{ij} = 1$; (ii) $w_{ij} = 0$ for treatments with $r_{ij} = 0$; and (iii) $w_{ij} = \hat{e}^1(X_{ij})$ for all controls, using predicted values from the fitted treatment group propensity score model.

3. Estimate $\hat{\tau}_{CACE-T}^{WLS}$ from a WLS model using the $w_{ij}$ weights that regresses $y_{ij}$ on an intercept, $T_j$, and $X_{ij}$ (Section 5.4 discusses standard error estimation).

We can also develop an IV estimator for the CACE-T estimand that does not rely on (A4a) but on a monotonicity condition. Recall that the SUTVA no-spillover condition implies an exclusion restriction on the never-takers ($\tau_{00} = 0$). Thus, invoking a monotonicity condition that treatments are at least as likely to receive services as controls implies that $\pi_{01} = 0$ (i.e., that there are no defiers), which yields the following IV estimator based on (4):

$$\hat{\tau}_{CACE-T}^{IV} = \frac{\hat{\tau}_{ITT}^{WLS}}{\hat{\pi}_{CACE-T}}. \tag{12}$$

Here, $\hat{\tau}_{ITT}^{WLS}$ is a standard regression-adjusted ITT estimator and $\hat{\pi}_{CACE-T} = \frac{1}{n^1}\sum_{j=1}^{m}\sum_{i=1}^{n_j} r_{ij}T_j$ is the proportion of treatment group members who received services.

---

[1] This WLS estimator does not have the doubly robust property [48], because the regression model without the correct weights estimates the ITT estimand, not the CACE-T estimand.

The plausibility of the monotonicity condition can be assessed by examining the study context and whether service receipt rates are considerably higher in the treatment than control group. With monotonicity, the CACE-T and CACE-TC estimands are the same.

### 5.2. Estimators for the $\tau_{CACE-TC}$ and $\tau_{11}$ Estimands

The above theory can be adapted to develop consistent estimators for $\tau_{CACE-TC}$ and $\tau_{11}$. For $\tau_{CACE-TC}$, we consider first the following estimator based on (4) that uses the SUTVA no-spillover condition ($\tau_{00} = 0$) and the (A4b) ignorability condition (but not monotonicity):

$$\hat{\tau}_{CACE-TC}^{WLS,1} = \frac{\hat{\tau}_{ITT}^{WLS}}{\hat{\pi}_{CACE-TC}}, \tag{13}$$

where $\hat{\pi}_{CACE-TC}$ is an IPW estimator for the proportion of the ITT population in the combined (1,1), (1,0), and (0,1) strata. This estimator is related to an IV estimator in that it scales up the ITT effects to pertain to treatment effects for compliers in the CACE-TC population. However, unlike the IV estimator in (12) for the CACE-T parameter, it does not rely on monotonicity ($\pi_{01} = 0$), but instead on (A4b) to obtain the $\hat{\pi}_{CACE-TC}$ estimators.

As shown in Appendix A (Result 2), two consistent IPW estimators for $\pi_{CACE-TC}$ are

$$\hat{\pi}_{CACE-TC}^1 = \frac{1}{n^1}\sum_{j=1}^{m}\sum_{i=1}^{n_j} T_j w_{ij}; \quad \hat{\pi}_{CACE-TC}^0 = \frac{1}{n^0}\sum_{j=1}^{m}\sum_{i=1}^{n_j}(1-T_j)w_{ij}, \tag{14}$$

where $w_{ij}$ are weights defined using

$$w_{ij} = r_{ij} + (1 - r_{ij})[T_j \hat{e}^0(\boldsymbol{X}_{ij}) + (1 - T_j)\hat{e}^1(\boldsymbol{X}_{ij})]. \tag{15}$$

Under (A4b) and (A6), $\hat{\pi}_{CACE-TC}^1$ and $\hat{\pi}_{CACE-TC}^0$ are asymptotically equivalent.

A second estimator for the $\tau_{CACE-TC}$ estimand is a standard weighted differences-in-means IPW estimator based on the weights defined in (15):

$$\hat{\tau}^2_{CACE-TC} = \bar{\bar{y}}^1_W - \bar{\bar{y}}^0_W = \frac{1}{w^1}\sum_{j=1}^{m}\sum_{i=1}^{n_j} w_{ij}T_j y_{ij} - \frac{1}{w^0}\sum_{j=1}^{m}\sum_{i=1}^{n_j} w_{ij}(1-T_j)y_{ij}, \qquad (16)$$

where $w^1 = \sum_{j=1}^{m}\sum_{i=1}^{n_j} w_{ij}T_j$ and $w^0 = \sum_{j=1}^{m}\sum_{i=1}^{n_j} w_{ij}(1-T_j)$. Appendix A (Result 3) discusses the consistency of $\hat{\tau}^2_{CACE-TC}$. The associated WLS estimator with covariates, $\hat{\tau}^{WLS,2}_{CACE-TC}$, is also consistent, which can be shown using the same methods as for the CACE-T estimator above.

Finally, an IPW estimator for the $\tau_{11}$ estimand relies on data for only the subsample of treatments and controls with $r_{ij} = 1$, where those with $r_{ij} = 0$ are excluded as they are not in the (1,1) population. The resulting IPW estimator, $\hat{\tau}_{11}$, can be obtained using (16) with $w_{ij} = r_{ij}[T_j\hat{e}^0(X_{ij}) + (1-T_j)\hat{e}^1(X_{ij})]$. Its consistency relies on the same conditions as for the CACE-TC estimator. The WLS estimator with covariates, $\hat{\tau}^{WLS}_{11}$, is also consistent.

### 5.3. Propensity Score Estimation

The considered IPW estimators require consistent estimators of the propensity scores, $e^1(X^1_{ij})$ and $e^0(X^0_{ij})$, where we now make it explicit that the $k^1$ covariates in the treatment group model can differ from the $k^0$ covariates in the control group model. Our analysis assumes a parametric logit model to estimate the propensity scores: $e^t(X^t_{ij}) = \exp(\alpha^t_0 + X^t_{ij}\alpha^t)/(1+\exp(\alpha^t_0 + X^t_{ij}\alpha^t))$ for $t \in \{1,0\}$, where $\alpha^t_0$ and $\alpha^t = (\alpha^t_1, \ldots, \alpha^t_{k^t})'$ are model parameters. We adopt the logit specification because it is commonly used in practice and facilitates standard error estimation of the treatment effects using stacked GEE methods that account for both the estimation error in the IPW weights and the clustered RCT design (see Section 5.4).

The covariates could include interaction and higher-order terms correlated with the primary outcomes, for example, informed by machine learning methods, [49-50]. The covariates cannot include cluster-specific fixed effects to capture potential cluster-level confounders (which yields inconsistent estimators but performs well in simulations [51]), because the weights for one research group are based on fitted models for the other group that are in different clusters.

*5.3.1. Specification Tests*

Standard specification tests for the propensity score models can be used to assess covariate balance across the actual and weighted service receipt groups, such as examining standardized differences in the covariates [52] and comparing densities of the estimated propensity scores using a variant of the Shaikh et al. method [53] (Appendix B). Confidence in the results could also be increased if measures of association between actual and predicted service receipt rates are high for the estimation samples (relative to random allocations). Though not a formal test of (A4b), it could also be useful to estimate separate logit models for subgroups that are strong predictors of service receipt and to examine the similarity of the coefficients across the models.

Another specification test is to gauge whether the mean weights—which pertain to the pertinent stratum complier population shares—are the same for the treatments and controls, which would occur if both logit models are specified correctly. For example, for the CACE-TC estimator in (14), $\hat{\pi}^1_{CACE-TC}$ and $\hat{\pi}^0_{CACE-TC}$ both estimate $(\pi_{11} + \pi_{10} + \pi_{01})$. One could also examine whether the CACE estimates conform to expectations. For instance, if the ITT effect is positive, we might expect the CACE-T effect to be larger than the CACE-TC and $\tau_{11}$ effects.

*5.3.2. Using Predictions on the Likelihood of Service Receipt*

From a data perspective, using IPW methods embedded in an RCT setting has the advantage that baseline data from a common source are often available for both the treatment and control groups. Service receipt decisions, however, are often complex and difficult to model in RCT settings, even when detailed baseline data are available. As discussed in [32], a simple baseline data collection design, which could improve the model predictions, is to collect information on the likely service receipt of individuals in the ITT sample, prior to random assignment if feasible, or if not, soon after entry into the study clusters. These predictions can be effective covariates in the propensity scoring models if they are accurate and available for most of the sample.

In some RCTs, there is an intake process in which staff complete intake forms (e.g., for hospital or clinic admissions or to determine an applicant's eligibility for a training program). In these cases, intake staff predictions can be obtained from supplemental forms to the intake forms, and the costs of obtaining these data should be low. In instances in which there is no formal application process (e.g., for school based RCTs), the prediction data can be obtained from forms or interviews completed by staff who have knowledge of individuals in the sample (e.g., teachers). An example of the successful use of this data collection design is in [54].

*5.4. Asymptotic Distributions and Standard Error Estimation*

The asymptotic distributions of the considered IPW estimators can be derived using GEE (M-estimation) theory [21]. In our context, the GEE approach involves the joint estimation of the logit and WLS outcome models, and extends the parallel GEE approach in [22-24] who consider non-clustered IPW designs that are not embedded within an RCT. Our approach reduces to non-clustered RCTs by treating clusters as individuals and setting $n_j = 1$.

To demonstrate the GEE approach, we focus on $\hat{\tau}_{CACE-TC}^{WLS,2}$, the regression-adjusted CACE-TC estimator in (16). Appendix C provides additional details for all estimators. To facilitate the analysis without changing the results, it is useful to parameterize the WLS regression model using $E[y_{ij}|T_j, X_{ij}] = T_j \mu_{CACE-TC}^1 + (1-T_j)\mu_{CACE-TC}^0 + X_{ij}\beta$ with no intercept, where $\mu_{CACE-TC}^1$ and $\mu_{CACE-TC}^0$ are mean potential outcomes in the CACE-TC population. Let $\boldsymbol{\xi} = (\mu_{CACE-TC}^1, \mu_{CACE-TC}^0, \boldsymbol{\beta}', \alpha_0^1, \boldsymbol{\alpha}^{1\prime}, \alpha_0^0, \boldsymbol{\alpha}^{0\prime})'$ denote the full vector of $(4 + k + k_1 + k_2)$ WLS and logit model parameters to be estimated.

We require several definitions pertaining to cluster-specific vectors. First, let $\mathbf{y}_j = (y_{1j}, y_{2j}, \ldots, y_{n_j j})'$ denote the $n_j x 1$ vector of person-level outcomes in cluster $j$, and similarly for $\mathbf{X}_j = (\mathbf{X}'_{1j}, \mathbf{X}'_{2j}, \ldots, \mathbf{X}'_{n_j j})'$ for the covariates in the WLS model, $\mathbf{X}^t_j = (\mathbf{X}^{t\prime}_{1j}, \mathbf{X}^{t\prime}_{2j}, \ldots, \mathbf{X}^{t\prime}_{n_j j})'$ for the covariates in the logit models for $t \in \{1,0\}$, $\mathbf{r}_j = (r_{1j}, r_{2j}, \ldots, r_{n_j j})'$ for the service receipt indicators, and $\mathbf{w}_j = (w_{1j}, w_{2j}, \ldots, w_{n_j j})'$ for the weights. Further, let $\mathbf{W}_j = diag(\mathbf{w}'_j)$ denote the $n_j x n_j$ diagonal matrix of weights, and define $\mathbf{1}_j = (1,1,\ldots,1)'$ as an $n_j x 1$ vector of 1s for the intercept terms. Finally, let $\mathbf{u}_j = (\mathbf{y}_j - T_j \mathbf{1}_j \mu_1 - (1-T_j)\mathbf{1}_j \mu_0 - \mathbf{X}_j \boldsymbol{\beta})$ denote WLS model residuals and $\boldsymbol{\eta}^t_j = (\mathbf{r}_j - \mathbf{e}^{*t}_j)$ denote logit model residuals, where $\mathbf{e}^{*t}_j(\mathbf{X}^t_j) = \exp(\mathbf{1}_j \alpha^t_0 + \mathbf{X}^t_j \boldsymbol{\alpha}^t)/(\mathbf{1}_j + \exp(\mathbf{1}_j \alpha^t_0 + \mathbf{X}^t_j \boldsymbol{\alpha}^t))$ is an $n_j x 1$ vector of individuals' true propensity scores for $t \in \{1,0\}$, with elementwise division of the numerator and denominator terms.

Using this notation, the GEE method for estimating $\hat{\boldsymbol{\xi}}$ solves the estimating equations, $\sum_{j=1}^{m} \boldsymbol{\psi}_j(\mathbf{y}_j, T_j, \mathbf{X}_j, \mathbf{X}^1_j, \mathbf{X}^0_j, \mathbf{W}_j, \mathbf{r}_j, \boldsymbol{\xi}) = \mathbf{0}$, where $\boldsymbol{\psi}_j(.)$ are cluster-specific score functions obtained from the first-order conditions (FOCs) for the WLS and logit objective functions:

$$\boldsymbol{\psi}_j(\mathbf{y}_j, T_j, \mathbf{X}_j, \mathbf{X}^1_j, \mathbf{X}^0_j, \mathbf{r}_j, \mathbf{W}_j, \boldsymbol{\xi}) = \begin{pmatrix} T_j \mathbf{1}'_j \mathbf{W}_j \mathbf{u}_j \\ (1-T_j)\mathbf{1}'_j \mathbf{W}_j \mathbf{u}_j \\ \mathbf{X}'_j \mathbf{W}_j \mathbf{u}_j \\ T_j \mathbf{1}'_j \boldsymbol{\eta}^1_j \\ T_j \mathbf{X}^{1\prime}_j \boldsymbol{\eta}^1_j \\ (1-T_j)\mathbf{1}'_j \boldsymbol{\eta}^0_j \\ (1-T_j)\mathbf{X}^{0\prime}_j \boldsymbol{\eta}^0_j \end{pmatrix}. \tag{17}$$

In this expression, the first three terms pertain to the FOCs for the $(2+k)$ parameters in the WLS model, while the last four terms pertain to the FOCs for the $(2+k_1+k_2)$ parameters in the treatment and control logit models. The $\boldsymbol{\psi}_j(.)$ terms are sums of variables across individuals in cluster $j$ (e.g., $T_j \mathbf{1}_j' \mathbf{W}_j \mathbf{u}_j = \sum_{i=1}^{n_j} T_j w_{ij} u_{ij}$). Under suitable regularity conditions [20,55], we have that $\sqrt{m}(\hat{\boldsymbol{\xi}} - \boldsymbol{\xi}) \xrightarrow{d} N(\mathbf{0}, \mathbf{V}(\boldsymbol{\xi}))$, where $\xrightarrow{d}$ denotes convergence in distribution and

$$\mathbf{V}(\boldsymbol{\xi}) = \boldsymbol{\Gamma}(\boldsymbol{\xi})^{-1} \boldsymbol{\Delta}(\boldsymbol{\xi}) [\boldsymbol{\Gamma}(\boldsymbol{\xi})^{-1}]' \tag{18}$$

is the asymptotic variance-covariance matrix, where $\boldsymbol{\Gamma}(\boldsymbol{\xi}) = \sum_j E(-\frac{d}{d\boldsymbol{\xi}'} \boldsymbol{\psi}_j(.))$ is the information matrix and $\boldsymbol{\Delta}(\boldsymbol{\xi}) = \sum_j E(\boldsymbol{\psi}_j(.) \boldsymbol{\psi}_j(.)')$. Here, expectations are taken with respect to outcomes and covariates in the CACE-TC complier superpopulation.

Note next that the estimated treatment effect, $\hat{\tau}_{CACE-TC} = (\hat{\mu}^1_{CACE-TC} - \hat{\mu}^0_{CACE-TC})$, can be obtained from $\hat{\boldsymbol{\xi}}$ using the relation, $\hat{\tau}_{CACE-TC} = \boldsymbol{\lambda}\hat{\boldsymbol{\xi}}$, where $\boldsymbol{\lambda} = (1, -1, 0, 0, \ldots, 0)$ is a $1 \times (4 + v + k_1 + k_2)$ row vector. Thus, $\hat{\tau}_{CACE-TC}$ is also asymptotically normal by the Delta method, and its asymptotic variance is $V(\hat{\tau}_{CACE-TC}) = \boldsymbol{\lambda} \mathbf{V}(\boldsymbol{\xi}) \boldsymbol{\lambda}'$. Consistent estimators for $V(\hat{\tau}_{CACE-TC})$ can be obtained using $\hat{V}(\hat{\tau}_{CACE-TC}) = \frac{1}{m} \boldsymbol{\lambda} \hat{\mathbf{V}}(\hat{\boldsymbol{\xi}}) \boldsymbol{\lambda}'$, where $\hat{\mathbf{V}}(\hat{\boldsymbol{\xi}})$ is calculated by inserting into (18) the plug-in estimators $\hat{\boldsymbol{\Gamma}}(\hat{\boldsymbol{\xi}}) = \frac{1}{m} \sum_{j=1}^{m} \left[-\frac{d}{d\boldsymbol{\xi}'} \hat{\boldsymbol{\psi}}_j(., \hat{\boldsymbol{\xi}})\right]$ and $\hat{\boldsymbol{\Delta}}(\hat{\boldsymbol{\xi}}) = \frac{1}{m} \sum_{j=1}^{m} \hat{\boldsymbol{\psi}}_j(., \hat{\boldsymbol{\xi}}) \hat{\boldsymbol{\psi}}_j(., \hat{\boldsymbol{\xi}})'$. This approach yields the empirical sandwich estimator. Appendix C displays the elements of $\boldsymbol{\psi}_j(.)$ and $\hat{\boldsymbol{\Gamma}}(\hat{\boldsymbol{\xi}})$ for all our considered IPW estimators.

While these variance calculations appear complex, they require only knowledge of $\boldsymbol{\psi}_j(.)$ and their first derivatives, which can be estimated using empirical counterparts by substituting $\hat{\boldsymbol{\xi}}$ for $\boldsymbol{\xi}$. No iteration is required. Appendix E displays sample R code for the calculations.

Hypothesis testing can be conducted using t-tests using the sandwich variance estimator and an estimator for the treatment effect, $\tau$. While the GEE estimator, $\hat{\tau}$, can be used for the analysis, the WLS estimator is typically used in practice. For clustered designs, it is customary to calculate the degrees of freedom ($df$) as the number of clusters ($m$) adjusted by the number of model parameters [46], although this issue is complex [56-57]. In our setting, using $df = (m - 4 - k - k^0 - k^1)$ for the CACE-TC estimator performs well in our simulations (see Section 6).

Hirano et al. [18] prove the surprising result (in a non-clustered setting) that the IPW estimator with nonparametrically estimated weights has a smaller variance than the IPW estimator using the "true" weights; thus, there are efficiency gains from using the estimated weights. A similar result is found in [23] for estimating average treatment effects using IPW weights from logit models without covariates, although [24] find that this result does not always hold for estimating treatment-on-the-treated effects. Obtaining parallel results for our more complex clustered IPW estimators is a topic for future research. However, in our simulations, we compare standard errors assuming the weights are known or estimated (see Section 6).

If the weights are assumed known (a common approach used in practice), one can use a simple consistent variance estimator developed in [40] for WLS estimators for clustered designs. This variance estimator is asymptotically equivalent to the GEE estimator but performs better in simulations when the number of clusters is relatively small (see Appendix C.3).

## 6. Simulations

We conducted simulations to examine the statistical performance of the considered IPW estimators. We examined biases of the estimators and the extent to which they achieve nominal confidence interval coverage and standard errors near true ones, under the assumed identification conditions. The simulations are based on plausible parameter values found in the literature. Our goal is not to examine the broader issue of how the estimators perform when their underlying assumptions are violated (which is a topic for future research), but to establish proof of concept.

To generate potential service receipt decisions for our simulations, we used a latent logistic binary choice framework (see Appendix D for simulation details):

$$R_{ij}^*(t) = \alpha_0^t + X_{ij1}\alpha_1^t + X_{ij2}\alpha_2^t + \varepsilon_{ij}^t,$$

$$R_{ij}(t) = 1 \text{ if } R_{ij}^*(t) \geq 0$$
$$R_{ij}(t) = 0 \text{ if } R_{ij}^*(t) < 0$$
(19)

for $t \in \{1,0\}$, where $R_{ij}^*(t)$ is a continuous net benefit (utility) measure underlying service receipt choices, and $\varepsilon_{ij}^1$ and $\varepsilon_{ij}^0$ are errors with logistic distributions (drawn independently to align with (A4b)). The same two covariates, $X_{ij1}$ and $X_{ij2}$, were included in the treatment and control models. They were generated independently using the relations: $X_{ij1} = u_{j1,X} + \varepsilon_{ij1,X}$ and $X_{ij2} = u_{j2,X} + \varepsilon_{ij2,X}$, where $u_{j1,X}$ and $u_{j2,X}$ are *iid* $N(0, \sigma_{uX}^2)$ cluster-specific components, and $\varepsilon_{ij1,X}$ and $\varepsilon_{ij2,X}$ are *iid* $N(0, \sigma_{\varepsilon X}^2)$ individual-level components. We selected the treatment group intercept, $\alpha_0^1$, to yield an expected treatment group service receipt rate of 50 or 70 percent, and we then selected the control group intercept, $\alpha_0^0$, so that the expected treatment-control service rate difference was 10 or 20 percentage points. The $\alpha_1^t$ and $\alpha_2^t$ parameters were selected to yield covariates with strong or moderate associations with service receipt (see Appendix D).

Next, we allocated the simulated sample to one of the four service receipt principal strata based on their simulated $(R_{ij}(1), R_{ij}(0))$ values, and then generated potential outcomes in principal stratum $(r, r')$ using the following model:

$$Y_{ij,rr'}(t) = \mu_{rr'}^t + X_{ij1} + X_{ij2} + u_{j,rr'} + T_j\theta_{j,rr'} + \epsilon_{ij,rr'}$$
(20)

for $(r, r', t) \in \{1,0\}$, where $u_{j,rr'}$, $\theta_{j,rr'}$ (which captures treatment effect heterogeneity), and $\epsilon_{ij,rr'}$ are each *iid* mean zero normal random errors. The stratum means, $\mu_{rr'}^t$, were selected to yield treatment effects—measured in effect size (standard deviation) units—that varied across the principal strata: $\tau_{11} = 0.20$, $\tau_{10} = 0.30$, $\tau_{01} = -0.10$, and $\tau_{00} = 0$ (to adhere to the SUTVA exclusion restriction of no spillover effects). Finally, we set standard deviations of the error terms and covariates to yield a regression $R^2$ value of 0.30 (so that the covariates are moderately correlated with the outcomes) and an intraclass correlation coefficient (clustering effect) of 0.10.

We ran 1,000 simulations for each specification, where the number of clusters was set to $m = 20$ or 80, and the number of individuals, $n_j$, was selected using random draws from a *Uniform*(40,80) distribution to allow for some size imbalance across clusters. At each simulation round, we randomly assigned clusters to either the treatment or control group using $p = 0.6$, and then generated $y_{ij}$ and $r_{ij}$ values based on the cluster treatment assignments using the relations in (1a) and (1b). Finally, for each round, we estimated the logit models and the WLS models with and without (i) covariates and (ii) standard error adjustments for estimation error in the weights.

Table 2 displays simulation results for several representative specifications where the service receipt rate is 70 percent for treatments and 50 percent for controls (Tables D.1 and D.2 display additional results). Several patterns emerge. First, biases of all estimators are small (less than 0.006 standard deviations in Table 2) even with $m = 20$ clusters. Second, estimated standard errors averaged across the simulations are similar to true ones (as measured by the standard deviation of the 1,000 estimated treatment effects), but tend to be slightly biased downward, especially for small $m$. These findings are consistent with other simulation evidence that the empirical sandwich estimator produces inflated Type 1 errors [46].

A third finding is that across specifications, adjusting for the estimation error in the weights yields similar standard errors as the unadjusted ones, with no pattern of differences (Table 2). Further, as suggested by theory, the use of covariates considerably improves precision without introducing bias. Precision does not noticeably differ across the estimators, except for the IV-related CACE-TC estimator, $\hat{\tau}_{CACE-TC}^{WLS,1}$, which has less power, as it scales up the aggregate ITT estimates rather than using outcome information directly linked to service recipients and nonrecipients. This finding highlights the power gains from using IPW versus IV estimators. Finally, all estimators achieve close to 95 percent confidence interval coverage based on t-distribution cutoff values and the degrees of freedom ($df$) values discussed in Section 5.4.

## 7. Empirical application

To demonstrate the use of the IPW estimators in practice and their broader application, we used data from a large-scale RCT of Early Head Start (EHS), a federal program for low-income pregnant women and families with infants and toddlers up to age three [25,62]. Funded by the U.S. Department of Health and Human Services, EHS is a two-generation program that fosters children's development while strengthening families. The program provides home-based services (through weekly home visits), center-based child care services (delivered in classroom settings), and family child care services, where specific services are tailored to the local context.

The study included 17 purposively selected EHS programs. Within each site, eligible families were randomly assigned to either a treatment group that was offered EHS services or a control group that was not. All families could receive other available services in their communities. The key study finding was that access to EHS services significantly increased children's cognitive and language development scores at age three [25].

The EHS study is a good case study for our methods because of the presence of service receipt effects, with balanced ITT samples [25] but unbalanced service receipt samples. During the study, 90 percent of treatments received home-based or child care services ("core" EHS services), compared to 38 percent of controls who received related services outside EHS. Further, the quality of the received services differed across the two research groups. For instance, measures of center quality tended to be higher for the EHS centers than the centers attended by the control group [25]. In addition, there were differences in the characteristics of service recipients across the research groups (Table E.1). For example, treatment recipients were more likely to be Hispanic and less likely to have a high school degree. There were also noticeable differences between service recipients and nonrecipients within each research group (Table E.1).

The ITT analysis conducted by Love et al. [25] treated the core services received by the control group as part of the study counterfactual. However, because of the presence of service receipt effects, the interpretation of these ITT estimates could be enhanced by estimating CACE effects for well-defined complier populations. Our analysis uses baseline data from study intake forms and outcome data (scale scores) from the Mental Development Index (MDI) from the Bayley Scales of Infant Development [58] administered to the focal children at age three. The sample includes 854 treatment and 756 control families with available MDI scores. The goal of our analysis is not to replicate evaluation findings, but to demonstrate our IPW methods.

We selected analysis covariates using Least Absolute Shrinkage and Selection Operator (lasso) procedures [59-61] applied to the covariates listed in Table E.1. Separate lasso models were estimated to identify predictive covariates for MDI scores (using the full sample) and service receipt (separately by research group). For each model, we first identified predictive main effects and then predictive two-way interactions among the predictive main effects. We used leave-one-out cross-validation to estimate the lasso regularization parameters. Let $\boldsymbol{X}_{MDI}$, $\boldsymbol{X}^1_{Logit}$, and $\boldsymbol{X}^0_{Logit}$ denote the lasso-selected covariates for the three models. The final logit model for research group $t \in \{1,0\}$ included the union of $\boldsymbol{X}^t_{Logit}$ and the main effects in $\boldsymbol{X}_{MDI}$. The final WLS model for CACE estimation included the union of all lasso-selected variables.

The results suggest that the IPW weights from the final logit models yield weighted samples with similar characteristics to those of actual service recipients in the other research group (Table E.2). The differences in mean covariate values between the samples are well below the 0.25 standard deviation threshold (and even the stricter 0.10 value) often used to assess adequate covariate balance [52]. Further, there is substantial overlap in the estimated propensity score distributions of the contrasted samples using the approach discussed in Appendix B (Figure E.1).

We find that estimates of the CACE-TC stratum shares in (14), $\hat{\pi}^1_{CACE-TC}$ and $\hat{\pi}^0_{CACE-TC}$, both yield the same value of 0.944, which provides additional support for the logit model specifications. Further, because $\hat{\pi}^1_{CACE-TC}$ and $\hat{\pi}^0_{CACE-TC}$ both estimate $(\pi_{11} + \pi_{10} + \pi_{01})$, it follows that $\hat{\pi}_{00} = 0.056$, or that 5.6 percent of the ITT population were never-takers (see Figure 1). In addition, because 90 percent of treatments and 38 percent of controls received core services, it follows that $\hat{\pi}_{11} = 0.346$, $\hat{\pi}_{10} = 0.564$, and $\hat{\pi}_{01} = 0.034$, suggesting a small number of defiers in the study population. These population shares are used below to help interpret the treatment effect estimates and to assess their plausibility.

The ITT estimate on MDI scores is 2.061 scale points with a standard error of 0.641 (Table 3), which translates into a gain of 0.15 standard deviations. The CACE-T and $\tau_{11}$ effects are 2.336 and 2.212 scale points, which, as expected, are larger than the ITT effect. Further, because the CACE-T effect is a weighted average of the $\tau_{11}$ and $\tau_{10}$ effects (with weights $\pi_{11}/(\pi_{11} + \pi_{10})$ and $\pi_{10}/(\pi_{11} + \pi_{10})$), we find that $\hat{\tau}_{10} = 2.412$. This yields the policy-relevant result that the ITT effect is due to *both* the effects of EHS on improving the quality of services for the always-takers (the $\tau_{11}$ effect) and increasing the take-up of services (the $\tau_{10}$ effect).

The two estimates for the CACE-TC estimand, $\hat{\tau}^{WLS,1}_{CACE-TC}$ and $\hat{\tau}^{WLS,2}_{CACE-TC}$, are nearly identical (2.183 and 2.186 scale points), with a somewhat larger standard error for the IV-related $\hat{\tau}^{WLS,1}_{CACE-TC}$ estimator, which is consistent with the simulation findings (Table 3). Further, because the CACE-TC estimand is a weighted average of the $\tau_{11}$, $\tau_{10}$, and $\tau_{01}$ effects, we find that $\hat{\tau}_{01} = -1.910$, suggesting that the treatment offer led to worse outcomes for the small number of defiers in the ITT population. These findings highlight that the CACE-TC estimand includes effects for both those who gain and lose from the treatment. In addition, using the ITT and CACE-TC estimates, we find insignificant effects for the never-takers ($\hat{\tau}_{00} = .0039$), supporting the exclusion restriction. Finally, similar to the simulation findings, we find few differences in the standard errors with and without adjustments for estimation error in the IPW weights.

## 8. Conclusions

This article developed causal estimands and simple IPW estimators for RCTs when the treatment affects the composition of service recipients in the ITT sample. Our methods apply to RCT settings where controls can access services related to the treatment. Using a principal stratification approach, we defined complier populations within the ITT sample based on potential service receipt choices in each research condition. This approach allows us to address research questions such as: "What are treatment effects for those likely to be served if the intervention was rolled out more broadly?" and "What are pure intervention effects that arise solely due to improved services for the always-takers?" The results from such analyses can be used to help interpret the ITT findings.

We developed IPW estimators that rely on ignorability conditions for parameter identification rather than distributional assumptions on potential outcomes invoked for other principal stratification estimators [19,41]. Our methods also provide an alternative to IV methods when standard monotonicity and exclusion restrictions might not hold. Our theoretical analysis established consistency of the IPW estimators. Further, because the IPW methods rely on predictive baseline covariates for the propensity score models of service receipt, we discussed a simple baseline data collection design that can substantially improve predictive power, where information on likely service receipt for each member of the ITT sample is obtained at study intake from pertinent site staff. We also discussed specification tests for the propensity score models, adapted to our RCT context.

While our analysis focused on clustered RCT designs with service receipt bias (the general case), our results also apply (reduce) to non-clustered RCTs, as demonstrated by our case study. We detailed a GEE approach where WLS is used to estimate treatment effects, using IPW weights from fitted logit propensity score models predicting service receipt, and where standard errors are obtained using the empirical sandwich estimator that accounts for estimation error in the weights. Our simulations show that the estimators have low bias and achieve nominal confidence interval coverage and standard errors near true ones, under the assumed identification conditions. The simulations also show, using plausible parameter values, that adjusting for estimation error in the weights is likely to have little effect on the estimated standard errors and significance levels. Finally, our case study illustrated how the IPW methods can address policy-relevant causal questions that can supplement and enrich the ITT findings.

Future research could extend the considered methods to clustered RCT settings with other forms of recruitment biases that can lead to unbalanced ITT samples. These biases could occur, for instance, in community-based trials where the treatment affects initial mobility into the treatment communities. Future extensions could also allow for multiple types of service receipt. Finally, additional simulations could be conducted to examine the statistical performance of the considered estimators to violations in their underlying assumptions.

**Data Availability Statement**. The EHS data for the empirical analysis was obtained under a restricted data use license agreement with ICPSR at the University of Michigan. Per license requirements, these data cannot be shared with journal readers. However, to the best of our knowledge, these data can be obtained from ICPSR, and the authors would be happy to provide their SAS programs used for the analysis. ICPSR bears no responsibility for use of the data or for interpretations or inferences based upon how it was used for this article.

**References**

1. Schochet PZ. Statistical power for random assignment evaluations of education programs. *Journal of Educational and Behavioral Statistics.* 2008; 33, 62-87.
2. Tipton E, Spybrook J, Fitzgerald KG, Wang Q, Davidson C. Toward a system of evidence for all: Current practices and future opportunities in 37 randomized trials. *Educational Researcher.* 2021; 50(3), 145-156.
3. Bland, JM. Cluster randomised trials in the medical literature: two bibliometric surveys, *BMC Medical Research Methodology*. 2004; 4, 21.
4. Johnson H, McNally S, Rolfe H, Ruiz-Valenzuela J, Savage R, Vousden J, Wood C. Teaching assistants, computers and classroom management. *Labour Economics*. 2019; 58, 21– 36.
5. What Works Clearinghouse. Standards Handbook, Version 4.1. U.S. Department of Education, Institute of Education Sciences, National Center for Education Evaluation and Regional Assistance, What Works Clearinghouse. 2020.
6. Campbell MK, Piaggio G, Elbourne DR, Altman DG. Consort 2010 Statement: Extension to cluster randomised trials, *BMJ*. 2012; 345 :e5661.
7. Bolzern J, Mnyama N, Bosanquet K, & Torgerson DJ. A review of cluster randomized trials found statistical evidence of selection bias. *Journal of Clinical Epidemiology*. 2018; 99, 106-112.
8. Brierley G, Brabyn S, Torgerson D, & Watson J. Bias in recruitment to cluster randomized trials: A review of recent publications. *J Eval Clin Pract*. 2012; 18(4), 878-886.
9. Donner A, Klar N. Pitfalls of and controversies in cluster randomization trials. *Am J Public Health*. 2004; 94, 416–22.
10. Eldridge S, Kerry S, Torgerson DJ. Bias in identifying and recruiting participants in cluster randomised trials: What can be done? *BMJ*. 2009; 339: b4006.
11. Eldridge S, Campbell M, Campbell M, Drahota-Towns A, Giraudeau B, Higgins J, Reeves B, Siegfried N. Revised Cochrane risk of bias tool for randomized trials (RoB 2.0): additional considerations for cluster-randomized trials. 2016; https://sites.google.com/site/riskofbiastool/welcome/rob-2-0-tool
12. Hahn S, Puffer S, Torgerson DJ. Methodological bias in cluster randomised trials. *BMC Med Res Methodology* . 2005; 5, 10.
13. Ivers NM, Taljaard M, Dixon S. Impact of CONSORT extension for cluster randomised trials on quality of reporting and study methodology: review of random sample of 300 trials, 2000-8. *BMJ*. 2011; 343.
14. Puffer S, Torgerson D, Watson J. Evidence for risk of bias in cluster randomised trials: Review of recent trials published in three general medical journals. *BMJ*. 2003; 327(7418), 785-789.

15. Turner EL, Li F, Gallis JA, Prague M, Murray DM. Review of recent methodological developments in group-randomized Trials: Part 1—design. *Am J Public Health*. 2017; 107(6), 907-915.
16. Horvitz DG, Thompson DJ. A generalization of sampling without replacement from a finite universe, *Journal of the American Statistical Association*. 1952; 47, 663–685.
17. Rosenbaum P. Model-based direct adjustment. *Journal of the American Statistical Association*. 1987; 82, 387-394.
18. Hirano K, Imbens G, Ridder G. Efficient estimation of average treatment effects using the estimated propensity score. *Econometrica*. 2003; 71(4): 1161-1189.
19. Frangakis CE, Rubin DB. (2002). Principal stratification in causal inference. *Biometrics*. 2002; 58, 20–29.
20. Angrist JD, Imbens GW, Rubin DB. Identification of causal effects using instrumental variables. *Journal of the American Statistical Association*. 1996; 91(434), 444-455.
21. Liang K, Zeger S. Longitudinal data analysis using generalized linear models. *Biometrika*. 1986; 73, 13-22.
22. Stefanski L, Boos D. The calculus of M-estimation. *The American Statistician*. 2002; 56(1), 29–38.
23. Lunceford JK, Davidian M. Stratification and weighting via the propensity score in estimation of causal treatment effects: A comparative study. *Statistics in Medicine*. 2004; 23(19), 2937–2960.
24. Reifeis S, Hudgens M. On variance of the treatment effect in the treated using inverse probability weighting. arXiv:2011.11874v1 [stat.ME]. 2020.
25. Love, JM, Kisker EM, Ross CM, Raikes H, Constantine JM, Boller K, Brooks-Gunn J, Chazan-Cohen R, Tarullo LB, Brady-Smith C, Fuligni AS, Schochet PZ, Paulsell DC, Vogel CA. The effectiveness of Early Head Start for 3-year-old children and their Parents: Lessons for policy and programs. *Developmental Psychology*. 2005; 41(6), 885-901.
26. Li F, Tian Z, Bobb J, Papadogeorgou G, Li F. Clarifying selection bias in cluster randomized trials: Estimands and estimation. Working Paper, Statistics Department, Duke University, Durham NC. 2021.
27. Schochet PZ. Student mobility, dosage, and principal stratification in clustered education RCTs of education interventions. *Journal of Educational and Behavioral Statistics*. 2013; 38(4), 323 354.
28. Cheng J, Small D. Bounds on causal effects in three-arm trials with noncompliance. *Journal of the Royal Statistical Society, Series B*. 2006; 68(5), 815–837.
29. Schochet PZ. The complier average causal effect parameter for multiarmed RCTs. *Evaluation Review*. 2020; 44(5-6), 410-436.
30. Leyrat C, Caille A, Donner A, Giraudeau B. Propensity scores used for analysis of cluster randomized trials with selection bias: A simulation study. *Statistics in Medicine*. 2013; 32(19), 3357-3372.

31. Peck LR. Subgroup analysis in social experiments measuring program impacts based on post-treatment choice. *American Journal of Evaluation*. 2003; 24, 157–187.
32. Schochet PZ, Burghardt JA. Using propensity scoring to estimate program-related subgroup impacts in experimental program evaluations. *Evaluation Review*. 2007; 31(2), 95-120.
33. Hong G, Deutsch J, Hill HD. Ratio-of-mediator-probability weighting for causal mediation analysis in the presence of treatment-by-mediator interaction. *Journal of Educational and Behavioral Statistics*. 2015; 40, 307–340.
34. Bein E, Deutsch J, Hong G, Porter K, Qin X, Yang C. Two-step estimation in RMPW analysis. *Statistics in Medicine*. 2018; 37(8), 1304-1324.
35. Leyrat C, Caille A, Foucher Y, Giraudeau B. Propensity score to detect baseline imbalance in cluster randomized trials: the role of the c-statistic. *BMC Med Res Methodol*. 2016; 16, 9. https://doi.org/10.1186/s12874-015-0100-4
36. Rubin DB. Which ifs have causal answers? Discussion of Holland's "Statistics and causal inference", *Journal of the American Statistical Association*. 1986; 81, 961-962.
37. Schochet PZ, Chiang H. Estimation and identification of the complier average causal effect parameter in education RCTs. *Journal of Educational and Behavioral Statistics*. 2011; 36(3), 307 345.
38. Kang H, Keele L. Spillover effects in cluster randomized trials with noncompliance. arXiv:1808.06418[stat.ME]. 2019.
39. Imbens G, Rubin D. Causal inference for statistics, social, and biomedical sciences: An introduction, Cambridge, UK: Cambridge University Press. 2015.
40. Schochet PZ, Pashley NE, Miratrix LW, Kautz T. Design-based ratio estimators and central limit theorems for clustered, blocked RCTs. *Journal of the American Statistical Association*. 2021. DOI: 10.1080/01621459.2021.1906685.
41. Zhang JL, Rubin DB, Mealli F. Likelihood-based analysis of causal effects of job-training programs using principal stratification. *Journal of the American Statistical Association*. 2009; 104(485), 166-176.
42. Feller A, Greif E, Ho N, Miratrix L, Pillai N. Weak separation in mixture models and implications for principal stratification. arXiv:1602.06595v2 [stat.ME]. 2019.
43. Rosenbaum P, Rubin DB. The central role of the propensity score in observational studies for causal effects. *Biometrika*. 1983; 70, 41–55.
44. Heckman J, Ichimura H, Smith J, Todd P. Characterizing selection bias using experimental data. *Econometrica*. 1998; 66 (5), 1017–1098.
45. Heckman JJ, Lalonde RJ, Smith JA. The economics and econometrics of active labor market programs. Handbook of Labor Economics, in: O. Ashenfelter & D. Card (ed.), Handbook of Labor Economics, edition 1, volume 3, chapter 31, pages 1865-2097, Elsevier. 1999.
46. Cameron AC, Miller DL. A practitioner's guide to cluster-robust inference, *Journal of Human Resources*. 2015; 50, 317-372.

47. Li X, Ding P. General forms of finite population central limit theorems with applications to causal inference. *Journal of the American Statistical Association.* 2017; 112: 1759-1769.
48. Robins JM, Rotnitzky A, Zhao LP. Estimation of regression coefficients when some regressors are not always observed, *Journal of the American Statistical Association*. 1994; 89, 846–866.
49. Athey S, Wager S. Estimating treatment effects with causal forests: An application. arXiv:1902.07409v1 [stat.ME]. 2019.
50. Belloni A, Chernozhukov V, Hansen C. Inference on treatment effects after selection among high-dimensional controls. *Review of Economic Studies*. 2014; 81, 608–650.
51. Yang S. Propensity score weighting for causal inference with clustered data. *Journal of Causal Inference*. 2018; 6(2), 2017-2027.
52. Stuart EA. Matching methods for causal inference: A review and a look forward. *Statistical Science*. 2010; 25(1), 1–21.
53. Shaikh AM, Simonsen M, Vytlacil EJ, Yildiz, N. A specification test for the propensity score using its distribution conditional on participation. *Journal of Econometrics*. 2009; 151(1), 33–46.
54. Schochet PZ, Burghardt JA, McConnell SM. Does Job Corps work? Impact findings from the National Job Corps Study. *American Economic Review*. 2008; 68(5), 1864-1886.
55. Hansen B, Lee S. Asymptotic theory for clustered samples. *Journal of Econometrics*. 2019; 210(2), 268-290.
56. Donald SG, Lang KL. Inference with difference-in-differences and other panel data, *Review of Economics and Statistics*. 2007; 89, 221–33.
57. Hedges L. Correcting a significance test for clustering. *Journal of Educational and Behavioral Statistics*. 2007; 32, 151-179.
58. Bayley, N. Bayley scales of infant development, 2nd ed.: Manual. New York: The Psychological Corporation. 1993.
59. Tibshirani R. Regression shrinkage and selection via the lasso. *Journal of the Royal Statistical Society, Series B*. 1996; 58: 267-288.
60. Hastie T, Tibshirani R, Friedman J. *The elements of statistical learning*. Springer, New York. 2009.
61. Hastie T, Tibshirani R., Wainwright M. Statistical learning with sparsity: The lasso and generalizations. Chapman & Hall/CRC. 2015.
62. United States Department of Health and Human Services. Administration for Children and Families. Early Head Start Research and Evaluation (EHSRE) Study, 1996-2010: [United States]. Inter-university Consortium for Political and Social Research [distributor], 2011-09-27. https://doi.org/10.3886/ICPSR03804.v.5.

**Table 1. Summary of key features of the considered IPW estimators**

| Estimand | Estimator | Equation number in text | Key identification assumption | IPW weights ($w_{ij}$) |
|---|---|---|---|---|
| CACE-T | $\hat{\tau}_{CACE-T}$; $\hat{\tau}_{CACE-T}^{WLS}$ | (10) and (11) | (A4a) | $w_{ij} = r_{ij}T_j + (1-T_j)\hat{e}^1(X_{ij}^1)$ |
| | $\hat{\tau}_{CACE-T}^{IV} = \hat{\tau}_{CACE-TC}^{IV}$ | (12) | Monotonicity | $w_{ij}^{ITT} = 1$ to estimate the ITT estimand; $w_{ij} = T_j r_{ij}$ to estimate $\hat{\pi}_{CACE-T}$ |
| CACE-TC | $\hat{\tau}_{CACE-TC}^1$; $\hat{\tau}_{CACE-TC}^{WLS,1}$ | (13) | (A4b) | $w_{ij}^{ITT} = 1$ to estimate the ITT estimand; $w_{ij} = T_j\left(r_{ij} + (1-r_{ij})\hat{e}^0(X_{ij}^0)\right)$ for $\hat{\pi}_{CACE-TC}^1$; $w_{ij} = (1-T_j)\left(r_{ij} + (1-r_{ij})\hat{e}^1(X_{ij}^1)\right)$ for $\hat{\pi}_{CACE-TC}^0$ |
| | $\hat{\tau}_{CACE-TC}^2$; $\hat{\tau}_{CACE-TC}^{WLS,2}$ | (16) | (A4b) | $w_{ij} = r_{ij} + (1-r_{ij})[T_j\hat{e}^0(X_{ij}^0) + (1-T_j)\hat{e}^1(X_{ij}^1)]$ |
| | $\hat{\tau}_{CACE-TC}^{IV}$ | (12) | Monotonicity | Same as $\hat{\tau}_{CACE-T}^{IV}$ above |
| $\tau_{11}$ | $\hat{\tau}_{11}$; $\hat{\tau}_{11}^{WLS}$ | (16) | (A4b) | $w_{ij} = r_{ij}[T_j\hat{e}^0(X_{ij}^0) + (1-T_j)\hat{e}^1(X_{ij}^1)]$ |

Notes. See text for variable definitions. IPW = inverse probability weighting; GEE = generalized estimating equations; ITT = intention-to-treat; WLS = weighted least squares.

**Table 2. Representative simulation results for the considered IPW estimators**

| Estimator and specification | True treatment effect[a] | Mean estimated treatment effect[a] | True standard error[b] | Mean estimated standard error[c] | Confidence interval coverage |
|---|---|---|---|---|---|
| **$m = 80$ Clusters** | | | | | |
| No covariates in the WLS models | | | | | |
| $\hat{\tau}_{CACE-T}$ | .249 | .248 | .099 | .096 / .097 | .939 |
| $\hat{\tau}^1_{CACE-TC}$ | .188 | .185 | .116 | .113 / .116 | .940 |
| $\hat{\tau}^2_{CACE-TC}$ | .188 | .186 | .098 | .096 / .096 | .940 |
| $\hat{\tau}_{11}$ | .200 | .198 | .101 | .098 / .099 | .940 |
| Two covariates in the WLS models | | | | | |
| $\hat{\tau}^{WLS}_{CACE-T}$ | .250 | .248 | .067 | .066 / .066 | .948 |
| $\hat{\tau}^{WLS,1}_{CACE-TC}$ | .188 | .186 | .077 | .078 / .077 | .949 |
| $\hat{\tau}^{WLS,2}_{CACE-TC}$ | .188 | .187 | .066 | .066 / .066 | .948 |
| $\hat{\tau}^{WLS}_{11}$ | .200 | .199 | .069 | .069 / .069 | .955 |
| **$m = 20$ Clusters** | | | | | |
| No covariates in the WLS models | | | | | |
| $\hat{\tau}_{CACE-T}$ | .251 | .245 | .187 | .182 / .183 | .949 |
| $\hat{\tau}^1_{CACE-TC}$ | .189 | .183 | .221 | .215 / .221 | .948 |
| $\hat{\tau}^2_{CACE-TC}$ | .189 | .184 | .187 | .182 / .182 | .951 |
| $\hat{\tau}_{11}$ | .201 | .195 | .192 | .186 / .187 | .950 |
| Two covariates in the WLS models | | | | | |
| $\hat{\tau}^{WLS}_{CACE-T}$ | .250 | .246 | .133 | .123 / .122 | .948 |
| $\hat{\tau}^{WLS,1}_{CACE-TC}$ | .188 | .185 | .156 | .150 / .156 | .956 |
| $\hat{\tau}^{WLS,2}_{CACE-TC}$ | .188 | .185 | .133 | .121 / .121 | .953 |
| $\hat{\tau}^{WLS}_{11}$ | .200 | .198 | .139 | .131 / .130 | .959 |

Notes. See text and Appendix D for simulation details. The calculations assume a mean of $n_j = 60$ individuals per cluster (randomly selected from a *Uniform*(40,80) distribution), a treatment group assignment rate of $p = .60$, a service receipt rate of 70 percent for treatments and 50 percent for controls, and that the covariates have a strong effect on the probability of service receipt for treatments and a moderate effect for controls. The figures are based on 1,000 simulations.

WLS = weighted least squares.

[a]Treatment effects are measured in effect size (standard deviation) units.
[b]True standard errors are measured as the standard deviation of the estimated treatment effects across simulations.
[c]The first standard error adjusts for estimation error in the weights, while the second standard error does not.

**Table 3. ITT and CACE estimates on MDI scores for the EHS Evaluation**

| Estimator | ITT or CACE treatment effect estimate (scale points) | Standard error | |
|---|---|---|---|
| | | Adjusted for estimation error in the IPW weights | Not adjusted for estimation error in the IPW weights |
| $\hat{\tau}_{ITT}^{WLS}$ | 2.061 | NA | .641 |
| $\hat{\tau}_{CACE-T}^{WLS}$ | 2.336 | .658 | .658 |
| $\hat{\tau}_{CACE-TC}^{WLS,1}$ | 2.183 | .704 | .679 |
| $\hat{\tau}_{CACE-TC}^{WLS,2}$ | 2.186 | .651 | .650 |
| $\hat{\tau}_{11}^{WLS}$ | 2.212 | .969 | .957 |

Notes: MDI = Mental Development Index scores measured at age three, measured in scale points [24]. The sample contains 854 treatments and 756 controls. The weights were constructed using the fitted logistic regression models. The ITT and CACE models included baseline covariates selected by lasso. See the text for a discussion of the estimators. NA = Not applicable; ITT = intention-to-treat; CACE = complier average causal effect. The standard deviation of the MDI scale scores is 13.6.

.

**Appendix A. Mathematical Proofs of Consistency Results for the IPW Estimators**

This appendix sketches proofs of the consistency of the considered IPW estimators as $m \to \infty$. For the proofs, we list the pertinent (A1)-(A6) conditions presented in the main text. The results also require suitable regularity conditions on population moments (not shown) that allow us to invoke weak laws of large numbers (see Liang & Zeger [21] and Stefanski & Boos [22]).

**Result 1.** Under Assumptions (A1)-(A3), (A4a), (A5), and (A6), the $\hat{\tau}_{CACE-T}$ estimator in (10) in the main text is a consistent estimator for $\tau_{CACE-T}$ as $m \to \infty$.

**Proof.** We first express $\hat{\tau}_{CACE-T}$ as follows:

$$\hat{\tau}_{CACE-T} = \bar{\bar{y}}_W^1 - \bar{\bar{y}}_W^0 = \frac{1}{w^1}\sum_{j=1}^{m}\sum_{i=1}^{n_j} r_{ij}T_j y_{ij} - \frac{1}{w^0}\sum_{j=1}^{m}\sum_{i=1}^{n_j} \hat{e}^1(X_{ij})(1-T_j)y_{ij}$$

$$= \frac{1}{w^1}\sum_{j=1}^{m}\sum_{i=1}^{n_j} R_{ij}(1)T_j Y_{ij}(1) - \frac{1}{w^0}\sum_{j=1}^{m}\sum_{i=1}^{n_j} \hat{e}^1(X_{ij})(1-T_j)Y_{ij}(0), \quad (A.1)$$

where the second equality is obtained by inserting the identities for $y_{ij}$ and $r_{ij}$ from (1a) and (1b) from the main text. If we now apply the weak law of large numbers for the treatment group mean (by invoking suitable regularity conditions), we find that

$$\bar{\bar{y}}_W^1 \xrightarrow{p} \frac{1}{E[R(1)T]} E[R(1)TY(1)], \quad (A.2)$$

where we have dropped the $i$ and $j$ subscripts and $t \in \{0,1\}$ superscripts on the covariates to reduce the already complex notation, and use the symbol $\xrightarrow{p}$ to denote convergence in probability. In this expression and elsewhere in this appendix, expectations are taken with respect to the potential outcomes and covariates in the complier superpopulation of interest. In (A.2), we have that $E[R(1)T] = (\pi_{11} + \pi_{10})p$, where $(\pi_{11} + \pi_{10}) = Pr(R(1) = 1|T = 1)$ is positive by Assumption (A5). Further, we have that

$$E[R(1)TY(1)] = E[Y(1)|R(1) = 1, T = 1](\pi_{11} + \pi_{10})p$$
$$= (\pi_{11}\mu_{11}^1 + \pi_{10}\mu_{10}^1)p. \tag{A.3}$$

Thus, we have that $\bar{\bar{y}}_W^1 \xrightarrow{p} \frac{1}{(\pi_{11}+\pi_{10})}(\pi_{11}\mu_{11}^1 + \pi_{10}\mu_{10}^1)$.

Next, note that the control group mean in (A.1), $\bar{\bar{y}}_W^0$, estimates $\frac{1}{E[e^1(X)(1-T)]}E[e^1(X)(1-T)Y(0)]$. Focusing on the numerator of this ratio and taking conditional expectations with respect to the covariate distribution, we find that

$$E[e^1(X)(1-T)Y(0)] = E_X\{E[e^1(X)(1-T)Y(0)|X]\}$$
$$= E_X\left\{e^1(X)\left(\frac{1}{e^1(X)}\right)E[R(1)Y(0)|T=0,X]\right\}(1-p) \tag{A.4}$$
$$= E[Y(0)|R(1) = 1](\pi_{11} + \pi_{10})(1-p).$$

In this expression, the second inequality relies on the independence of $R(1)$ and $Y(0)$ conditional on the covariates (which (A4a) assumes) and on the relation in (7) from the main text that $Pr(R(1) = 1|T = 0, X) = e^1(X)$. Similarly, we have that $E[e^1(X)(1-T)] = (\pi_{11} + \pi_{10})(1-p)$. Thus, $\bar{\bar{y}}_W^0 \xrightarrow{p} E[Y(0)|R(1) = 1] = \frac{1}{(\pi_{11}+\pi_{10})}(\pi_{11}\mu_{11}^0 + \pi_{10}\mu_{10}^0)$, and it then follows that $\hat{\tau}_{CACE-T} \xrightarrow{p} \tau_{CACE-T}$.

**Result 2.** Under (A1)-(A3), (A4b), (A5), and (A6), the $\hat{\pi}_{CACE-TC}^1$ and $\hat{\pi}_{CACE-TC}^0$ estimators in (14) in the main text are consistent estimators for $\pi_{CACE-TC}$.

**Proof.** We prove this result for $\hat{\pi}_{CACE-TC}^1$, as similar arguments apply to $\hat{\pi}_{CACE-TC}^0$. We first split $\hat{\pi}_{CACE-TC}^1$ into separate terms for the $r_{ij} = 1$ and $r_{ij} = 0$ service groups. This approach yields the following expression:

$$\hat{\pi}_{CACE-TC}^1 = \frac{1}{np}\sum_{j=1}^{m}\sum_{i=1}^{n_j}R_{ij}(1)T_j + \frac{1}{np}\sum_{j=1}^{m}\sum_{i=1}^{n_j}\hat{e}^0(X_{ij})(1-R_{ij}(1))T_j. \tag{A.5}$$

If we now apply the weak law of large numbers to each term in (A.5), we find that

$$\hat{\pi}^1_{CACE-TC} \xrightarrow{p} \frac{1}{p}\{E[R(1)T] + E[e^0(X)(1 - R(1))T]\}. \tag{A.6}$$

The first expectation in (A.6) equals $(\pi_{11} + \pi_{10}) > 0$. For the second term, we can use the law of iterated expectations, conditioning over the distribution of covariate values, to show that

$$\frac{1}{p}E[e^0(X)(1 - R(1))T] = \frac{1}{p}E_X\{E[e^0(X)(1 - R(1))T|X]\}$$

$$= \frac{1}{p}E_X\{E[e^0(X)|R(1) = 0, T = 1, X]\}(\pi_{01} + \pi_{00}) \tag{A.7}$$

$$= E_X\left\{e^0(X)\left(\frac{1}{e^0(X)}\right)E[R(0)|R(1) = 0, T = 1, X]\right\}(\pi_{01} + \pi_{00})$$

$$= E[R(0)|R(1) = 0, T = 1](\pi_{01} + \pi_{00}) = \pi_{01},$$

where $(\pi_{01} + \pi_{00}) = Pr(R(1) = 0|T = 1) > 0$. For the second equality, we use the key result, $Pr(R(0) = 1|R(1) = 0, T = 1, X) = e^0(X)$, which follows from (A4b) and (A2). Thus, if we insert all the pieces into (A.6), we find that $\hat{\pi}^1_{CACE-TC} \xrightarrow{p} \pi_{CACE-TC} = (\pi_{11} + \pi_{10} + \pi_{01})$.

**Result 3.** Under (A1)-(A3), (A4b), (A5), and (A6), the $\hat{\tau}^2_{CACE-TC}$ estimator in (16) in the main text is a consistent estimator for $\tau_{CACE-TC}$ as $m \to \infty$.

**Proof.** Similar to the proof of Result 2, we first split the summations in (16) into separate terms for the $r_{ij} = 1$ and $r_{ij} = 0$ service groups. This yields the following expression for the treatment group mean:

$$\bar{\bar{y}}^1_W = \frac{1}{w^1}\sum_{j=1}^{m}\sum_{i=1}^{n_j}R_{ij}(1)T_jY_{ij}(1) + \frac{1}{w^1}\sum_{j=1}^{m}\sum_{i=1}^{n_j}\hat{e}^0(X_{ij})(1 - R_{ij}(1))T_jY_{ij}(1). \tag{A.8}$$

A similar expression holds for the control group mean. Note next that

$$\bar{\bar{y}}^1_W \xrightarrow{p} \frac{1}{E[w^1]}\{E[R(1)T(Y(1)] + E[e^0(X)(1 - R(1))TY(1)]\}. \tag{A.9}$$

The first term in the numerator of (A.9) can be expressed as $E[Y(1)|R(1) = 1, T = 1](\pi_{11} + \pi_{10})p = (\pi_{11}\mu_{11}^1 + \pi_{10}\mu_{10}^1)p$, where $(\pi_{11} + \pi_{10}) = Pr(R(1) = 1|T = 1)$ is positive by (A5). For the second numerator term, we can use the law of iterated expectations, conditioning over the distribution of covariate values, to show that

$$E[e^0(X)(1 - R(1))TY(1)] = E_X\{E[e^0(X)(1 - R(1))TY(1)|X]\}$$
$$= E_X\{E[e^0(X)Y(1)|R(1) = 0, T = 1, X]\}(\pi_{01} + \pi_{00})p$$
$$= E_X\left\{e^0(X)\left(\frac{1}{e^0(X)}\right)E[R(0)Y(1)|R(1) = 0, T = 1, X]\right\}(\pi_{01} + \pi_{00})p$$
$$= E[R(0)Y(1)|R(1) = 0, T = 1](\pi_{01} + \pi_{00})p = \pi_{01}\mu_{01}^1 p,$$
(A.10)

where $(\pi_{01} + \pi_{00}) = Pr(R(1) = 0|T = 1) > 0$. For the second equality, we use the key result, $Pr(R(0) = 1|R(1) = 0, T = 1, X) = e^0(X)$, which follows from (A4b) and (A2). Finally, similar methods show that $E[w^1] = E[R(1)T] + E[e^0(X)(1 - R(1))T] = (\pi_{11} + \pi_{10})p + \pi_{01}p$. Thus, if we insert all the pieces into (A.8), we find that $\bar{\bar{y}}_W^1 \xrightarrow{p} \frac{1}{\pi_{CACE-TC}}(\pi_{11}\mu_{11}^1 + \pi_{10}\mu_{10}^1 + \pi_{01}\mu_{01}^1)$. Applying similar methods for the control group, we find that $\bar{\bar{y}}_W^0 \xrightarrow{p} \frac{1}{\pi_{CACE-TC}}(\pi_{11}\mu_{11}^0 + \pi_{10}\mu_{10}^0 + \pi_{01}\mu_{01}^0)$. Thus, $\hat{\tau}_{CACE-TC} \xrightarrow{p} \tau_{CACE-TC}$.

**Appendix B. The Shaikh et al. [53] Specification Test Adapted to the IPW Estimators**

We can adapt to our setting the specification test in Shaikh et al. [53] that compares the densities of propensity scores from parametric models for treated and untreated populations. Although Shaikh et al. [53] consider matching rather than IPW estimators, their methods also apply to our IPW setting for propensity score models based on service recipients and nonrecipients in each research group as shown by the following result:

**Result.** Let $f^1(q)$ and $f^0(q)$ denote the densities (with respect to Lebesgue measure) of $e^1(X)$ and $e^0(X)$ for $0 < q < 1$. Further let $f^1_{R(1)=t}(q)$ denote the density of the propensity scores for the treatment group conditional on $R(1) = t$ for $t \in \{1, 0\}$, and similarly for $f^0_{R(0)=t}(q)$ for the control group. Assume that $Pr(R(1) = 1 | e^1(X) = q) > 0$ for all $q$ in the support of $e^1(X)$ for the treatment group, and similarly for the control group. Then

$$f^1_{R(1)=1}(q) = \omega^1 \frac{q}{(1-q)} f^1_{R(1)=0}(q); \quad f^0_{R(0)=1}(q) = \omega^0 \frac{q}{(1-q)} f^0_{R(0)=0}(q), \quad (B.1)$$

where $\omega^1 = \frac{Pr(R(1)=0)}{Pr(R(1)=1)}$ and $\omega^0 = \frac{Pr(R(0)=0)}{Pr(R(0)=1)}$ are population ratios based on service receipt rates.

Before proving this result, note that in our setting, the relations in (B.1) will hold if the propensity score logit models predicting service receipt for treatments and controls are specified correctly. Thus, graphing empirical estimates of $f^1_{R(1)=1}(q)$ using the estimated propensity scores from the fitted logit models against empirical estimates of $\omega^1 \frac{q}{(1-q)} f^1_{R(1)=0}(q)$ can be used to assess model misspecification for the treatment group logit model, and similarly for the control group logit model. The $\omega^1$ and $\omega^0$ ratios can be estimated observed service receipt rates. Shaikh et al. [53] also discuss an asymptotically normal test statistic for the restrictions in (B.1) that applies to our setting as well.

**Proof.** Following Shaikh et al. [53], we can establish the results in (B.1) by applying Bayes Theorem to each conditional density. For the $f^1_{R(1)=1}(q)$ density, Bayes Theorem implies that

$$f^1_{R(1)=1}(q) Pr(R(1) = 1) = Pr(e^1(X) = q | R(1) = 1) Pr(R(1) = 1)$$
$$= Pr(R(1) = 1 | e^1(X) = q) f^1(q) = q f^1(q). \quad (B.2)$$

The last equality follows from the law of iterated expectations because

$$E[R(1)|e^1(X) = q] = E_{e^1(X)}\{E[R(1)|e^1(X), X]|e^1(X) = q\}$$
$$= E_{e^1(X)}\{E[R(1)|X]|e^1(X) = q\} = q.$$
(B.3)

We have that $qf^1(q) > 0$ by our support assumption. Similarly, we find that $f^1_{R(1)=0}(q)Pr(R(1) = 0) = (1-q)f^1(q) > 0$. Thus, calculating the ratio, $f^1_{R(1)=1}(q)/f^1_{R(1)=0}(q)$, yields the first result in (B.1) for the treatment group, and a similar argument leads to the second result in (B.1) for the control group.

**Appendix C. Additional Details for the GEE Estimators**

In this appendix, we first discuss the GEE approach for the IV-related estimators in (12) and (13) as it differs somewhat from the differences-in-means IPW estimators discussed in the main text. We then discuss the elements of the $\boldsymbol{\xi}$ parameter vectors and cluster-specific score functions, $\boldsymbol{\psi}_j(.)$, for all estimators, where we focus on the general case where covariates are included in the WLS regression models. Finally, we discuss the variance estimator in Schochet et al. [40] that can be used if the weights are assumed known.

*C.1. GEE Approach for the IV-Related Estimators*

The IV-related CACE-TC estimator in (13), $\hat{\tau}^{WLS,1}_{CACE-TC} = \frac{\hat{\tau}^{WLS}_{ITT}}{\hat{\pi}_{CACE-TC}}$, is a ratio estimator with estimation error in both the numerator and denominator terms. Suppose $\hat{\pi}_{CACE-TC}$ is estimated using $\hat{\pi}^1_{CACE-TC} = \frac{1}{n^1}\sum_{j=1}^{m}\sum_{i=1}^{n_j} T_j w_{ij}$ with the weights in (15) (similar arguments apply using $\hat{\pi}^0_{CACE-TC}$). The GEE parameter vector is $\boldsymbol{\xi} = (\mu^1, \mu^0, \boldsymbol{\beta}', \alpha^0_0, \boldsymbol{\alpha}^{0'}, \pi_{CACE-TC}, \tau_{CACE-TC})'$, where $\tau_{CACE-TC} = \frac{(\mu^1 - \mu^0)}{\pi_{CACE-TC}}$ is the estimand of interest. The score function is

$$\boldsymbol{\psi}_j(\boldsymbol{y}_j, T_j, \boldsymbol{X}_j, \boldsymbol{X}_j^0, \boldsymbol{r}_j, \boldsymbol{W}_j, \boldsymbol{\xi}) = \begin{pmatrix} T_j \mathbf{1}_j' \boldsymbol{u}_j \\ (1-T_j) \mathbf{1}_j' \boldsymbol{u}_j \\ \boldsymbol{X}_j' \boldsymbol{u}_j \\ (1-T_j) \mathbf{1}_j' \boldsymbol{\eta}_j^0 \\ (1-T_j) \boldsymbol{X}_j^{0'} \boldsymbol{\eta}_j^0 \\ (T_j \mathbf{1}_j' \mathbf{1}_j) \pi_{CACE-TC} - T_j \mathbf{1}_j' \boldsymbol{w}_j \\ \tau_{ITT,X} - \tau_{CACE-TC} \pi_{CACE-TC} \end{pmatrix}, \tag{C.1}$$

where $\tau_{ITT,X} = [(\mu_1 - \mu_0) - (\bar{\bar{X}}_W^1 - \bar{\bar{X}}_W^0)\boldsymbol{\beta}]$ is the regressed-adjusted ITT estimator, and the last two rows pertain to estimation of the $\pi_{CACE-TC}$ population share and the ratio estimator.

The resulting GEE estimator, $\hat{\boldsymbol{\xi}}$, is asymptotically normal with variance, $V(\boldsymbol{\xi})$, as defined in (18) in the main text. Thus, the ratio estimator, $\hat{\tau}_{CACE-TC}^{WLS,1}$, has asymptotic variance, $V(\hat{\tau}_{CACE-TC}^{1,WLS}) = V(\boldsymbol{\xi})_{(l,l)}$, where $l = (5 + k + k^0)$ signifies the last entry of $V(\boldsymbol{\xi})$. This variance can be estimated using the empirical sandwich estimator. Weights are not required for the WLS analysis to estimate the ITT (numerator) component of the ratio estimator. A similar GEE approach can be used for the IV estimator for the CACE-T and CACE-TC estimands, $\hat{\tau}_{CACE-T}^{IV} = \hat{\tau}_{CACE-TC}^{IV}$, in (12). Note that if we ignore the estimation error in the $\hat{e}_{ij}^0(\boldsymbol{X}_{ij}^0)$ weights, we can use the variance estimator in Schochet et al. [40], dividing by $(\hat{\pi}_{CACE-TC}^1)^2$ and setting $\hat{w}_{ij} = 1$.

*C.2. The GEE Parameter Vectors and Score Functions*

The elements of the GEE parameter vectors, $\boldsymbol{\xi}$, for the considered estimators and the associated cluster-specific score functions, $\boldsymbol{\psi}_j(.)$, are shown in Table C.1. The score functions are obtained from the first-order conditions for the WLS and logit objective functions that minimize the weighted sums of squared residuals across the full sample.

**Table C.1. Parameter vectors ($\xi$) and score functions ($\psi_j$) for the considered estimators**

| $\hat{\tau}^{WLS}_{CACE-T}$ | $\hat{\tau}^{1,WLS}_{CACE-TC}$ [a] | $\hat{\tau}^{2,WLS}_{CACE-TC}$ | $\hat{\tau}^{WLS}_{11}$ |
|---|---|---|---|
| Parameter vectors | | | |
| $\begin{pmatrix} \mu^1_{CACE-T} \\ \mu^0_{CACE-T} \\ \boldsymbol{\beta} \\ \alpha^1_0 \\ \boldsymbol{\alpha}^1 \end{pmatrix}$ | $\begin{pmatrix} \mu^1 \\ \mu^0 \\ \boldsymbol{\beta} \\ \alpha^0_0 \\ \boldsymbol{\alpha}^0 \\ \pi_{CACE-TC} \\ \tau_{CACE-TC} \end{pmatrix}$ | $\begin{pmatrix} \mu^1_{CACE-T} \\ \mu^0_{CACE-T} \\ \boldsymbol{\beta} \\ \alpha^1_0 \\ \boldsymbol{\alpha}^1 \\ \alpha^0_0 \\ \boldsymbol{\alpha}^0 \end{pmatrix}$ | $\begin{pmatrix} \mu^1_{\tau(1,1)} \\ \mu^0_{\tau(1,1)} \\ \boldsymbol{\beta} \\ \alpha^1_0 \\ \boldsymbol{\alpha}^1 \\ \alpha^0_0 \\ \boldsymbol{\alpha}^0 \end{pmatrix}$ |
| Score functions | | | |
| $\begin{pmatrix} T_j \mathbf{1}'_j W_j u_j \\ (1-T_j)\mathbf{1}'_j W_j u_j \\ X'_j W_j u_j \\ T_j \mathbf{1}'_j \eta^1_j \\ T_j X^{1'}_j \eta^1_j \end{pmatrix}$ | $\begin{pmatrix} T_j \mathbf{1}'_j u_j \\ (1-T_j)\mathbf{1}'_j u_j \\ X'_j u_j \\ (1-T_j)\mathbf{1}'_j \eta^0_j \\ (1-T_j)X^{0'}_j \eta^0_j \\ (T_j \mathbf{1}'_j \mathbf{1}_j)\pi_{CACE-TC} - T_j \mathbf{1}'_j w_j \\ \tau_{ITT,X} - \tau_{CACE-TC}\pi_{CACE-TC} \end{pmatrix}$ | $\begin{pmatrix} T_j \mathbf{1}'_j W_j u_j \\ (1-T_j)\mathbf{1}'_j W_j u_j \\ X'_j W_j u_j \\ T_j \mathbf{1}'_j \eta^1_j \\ T_j X^{1'}_j \eta^1_j \\ (1-T_j)\mathbf{1}'_j \eta^0_j \\ (1-T_j)X^{0'}_j \eta^0_j \end{pmatrix}$ | $\begin{pmatrix} T_j \mathbf{1}'_j W_j u_j \\ (1-T_j)\mathbf{1}'_j W_j u_j \\ X'_j W_j u_j \\ T_j \mathbf{1}'_j \eta^1_j \\ T_j X^{1'}_j \eta^1_j \\ (1-T_j)\mathbf{1}'_j \eta^0_j \\ (1-T_j)X^{0'}_j \eta^0_j \end{pmatrix}$ |

Note: The parameter vectors and score functions for the WLS models without covariates exclude the third rows.
[a] Assumes $\hat{\pi}^1_{CACE-TC}$ is used for estimation. A similar score function exists using $\hat{\pi}^0_{CACE-TC}$ instead.

These score functions can be used to obtain the GEE empirical sandwich estimators using

$\hat{V}(\hat{\xi}) = \frac{1}{m}\hat{\Gamma}(\hat{\xi})^{-1}\hat{\Delta}(\hat{\xi})[\hat{\Gamma}(\hat{\xi})^{-1}]'$, where $\hat{\Gamma}(\hat{\xi}) = \frac{1}{m}\sum_{j=1}^{m}\hat{\Gamma}_j(\hat{\xi})$, $\hat{\Gamma}_j(\hat{\xi}) = \left[-\frac{d}{d\xi'}\hat{\psi}_j(.,\hat{\xi})\right]$, and

$\hat{\Delta}(\hat{\xi}) = \frac{1}{m}\sum_{j=1}^{m}\hat{\psi}_j(.,\hat{\xi})\hat{\psi}_j(.,\hat{\xi})'$. It is straightforward to calculate $\hat{\Delta}(\hat{\xi})$ using $\hat{\psi}_j(.,\hat{\xi})$.

Calculating $\hat{\Gamma}(\hat{\xi})$ is also straightforward but is more complex as it requires first derivatives. To illustrate the calculations, below, we provide the elements of $\hat{\Gamma}_j(\hat{\xi})$ for the CACE-TC estimator in (16) in the main text:

$$\hat{\Gamma}_{j,CACE-TC}(\hat{\xi}) = \begin{pmatrix} A_{(k+2)x(k+2)} & B_{(k+2)x(k^1+k^0+2)} \\ \mathbf{0}_{(k^1+k^0+2)x(k+2)} & C_{(k^1+k^0+2)x(k^1+k^0+2)} \end{pmatrix}, \quad (C.2)$$

where

$$A = \begin{pmatrix} \sum_{i=1}^{n_j} T_j \widehat{w}_{ij} & 0 & \sum_{i=1}^{n_j} T_j \widehat{w}_{ij} X_{ij} \\ 0 & \sum_{i=1}^{n_j} (1-T_j) \widehat{w}_{ij} & \sum_{i=1}^{n_j} (1-T_j) \widehat{w}_{ij} X_{ij} \\ \sum_{i=1}^{n_j} X'_{ij} T_j \widehat{w}_{ij} & \sum_{i=1}^{n_j} X'_{ij} (1-T_j) \widehat{w}_{ij} & \sum_{i=1}^{n_j} \widehat{w}_{ij} X'_{ij} X_{ij} \end{pmatrix},$$

$$B = \begin{pmatrix} \mathbf{0}_{1 \times (k^1+1)} & \sum_{i=1}^{n_j} -T_j(1-r_{ij}) \hat{u}_{ij} \hat{e}^0_{ij}(1-\hat{e}^0_{ij})(1 \ X^0_{ij}) \\ \sum_{i=1}^{n_j} -(1-T_j)(1-r_{ij}) \hat{u}_{ij} \hat{e}^1_{ij}(1-\hat{e}^1_{ij})(1 \ X^1_{ij}) & \mathbf{0}_{1 \times (k^0+1)} \\ \sum_{i=1}^{n_j} -X'_{ij}(1-T_j)(1-r_{ij}) \hat{u}_{ij} \hat{e}^1_{ij}(1-\hat{e}^1_{ij})(1 \ X^1_{ij}) & \sum_{i=1}^{n_j} -X'_{ij} T_j (1-r_{ij}) \hat{u}_{ij} \hat{e}^0_{ij}(1-\hat{e}^0_{ij})(1 \ X^0_{ij}) \end{pmatrix},$$

and

$$C = \begin{pmatrix} \sum_{i=1}^{n_j} T_j \hat{e}^1_{ij}(1-\hat{e}^1_{ij})(1 \ X^1_{ij}) & \mathbf{0}_{1 \times (k^0+1)} \\ \sum_{i=1}^{n_j} X^{1\prime}_{ij} T_j \hat{e}^1_{ij}(1-\hat{e}^1_{ij})(1 \ X^1_{ij}) & \mathbf{0}_{k^1 \times (k^0+1)} \\ \mathbf{0}_{1 \times (k^1+1)} & \sum_{i=1}^{n_j} (1-T_j) \hat{e}^0_{ij}(1-\hat{e}^0_{ij})(1 \ X^0_{ij}) \\ \mathbf{0}_{k^0 \times (k^1+1)} & \sum_{i=1}^{n_j} X^{0\prime}_{ij}(1-T_j) \hat{e}^0_{ij}(1-\hat{e}^0_{ij})(1 \ X^0_{ij}) \end{pmatrix}.$$

In this expression, $A$ pertains to the WLS model parameters, $B$ pertains to the logit model parameters, and $C$ pertains to the logit model parameters for the IPW weights in the WLS model.

The $\hat{\mathbf{\Gamma}}_j(\hat{\xi})$ matrix in (C.2) can also be used for the other IPW estimators with their respective weights with a few changes. For the CACE-T estimator, $\hat{\tau}_{CACE-T}$, the changes are: (i) omitting the right panel terms (columns) in $B$, (ii) deleting the $(1-r_{ij})$ terms in the left panel in $B$, and (iii) only including the $\sum_{i=1}^{n_j} T_j \hat{e}^1_{ij}(1-\hat{e}^1_{ij})(1 \ X^1_{ij})$ and $\sum_{i=1}^{n_j} X^{1\prime}_{ij} T_j \hat{e}^1_{ij}(1-\hat{e}^1_{ij})(1 \ X^1_{ij})$ terms in $C$ (because the control group logit model is not required). For the $\hat{\tau}_{11}$ estimator, the only change is to replace the $(1 - r_{ij})$ terms in $\boldsymbol{B}$ with $r_{ij}$.

The $\hat{\boldsymbol{\Gamma}}_j(\hat{\boldsymbol{\xi}})$ matrix for the IV-related CACE-TC ratio estimator in (13) of the main text differs somewhat from (C.2) due to the inclusion of the ratio estimator term. If we use $\hat{\pi}^1_{CACE-TC}$ to estimate the CACE-TC population share, we have that

$$\hat{\boldsymbol{\Gamma}}^{WLS,1}_{CACE-TC}(\hat{\boldsymbol{\xi}}) = \begin{pmatrix} \boldsymbol{A}_{(k+2)x(k+2)} & \boldsymbol{0}_{(k+2)x(k^0+3)} \\ \boldsymbol{B}_{(k^0+3)x(k+2)} & \boldsymbol{C}_{(k^0+2)x(k^0+3)} \end{pmatrix}, \tag{C.3}$$

where $\boldsymbol{A}$ is the same as in (C.2) except without weights, $\boldsymbol{B}$ is now a matrix of 0s except the last row is $(-1 \quad 1 \quad (\overline{\overline{\boldsymbol{X}}}^1_W - \overline{\overline{\boldsymbol{X}}}^0_W))$ and

$$\boldsymbol{C} = \begin{pmatrix} \sum_{i=1}^{n_j}(1 - T_j)\hat{e}^0_{ij}(1 - \hat{e}^0_{ij})(1 \quad \boldsymbol{X}^0_{ij}) & 0 & 0 \\ \sum_{i=1}^{n_j} \boldsymbol{X}^{0\prime}_{ij}(1 - T_j)\hat{e}^0_{ij}(1 - \hat{e}^0_{ij})(1 \quad \boldsymbol{X}^0_{ij}) & 0 & 0 \\ \sum_{i=1}^{n_j} T_j(1 - r_{ij})\hat{e}^0_{ij}(1 - \hat{e}^0_{ij})(1 \quad \boldsymbol{X}^0_{ij}) & -T_j n_j & 0 \\ \boldsymbol{0}_{1x(k^0+1)} & \hat{\tau}^1_{CACE-TC} & \hat{\pi}^1_{CACE-TC} \end{pmatrix}.$$

A similar expression applies if we instead use $\hat{\pi}^0_{CACE-TC}$ to estimate the denominator term.

Finally, the score function for the CACE-T parameter in (12), $\hat{\tau}^{IV}_{CACE-T}$, is

$$\boldsymbol{\psi}_j(.) = \begin{pmatrix} T_j \boldsymbol{1}'_j \boldsymbol{u}_j \\ (1 - T_j)\boldsymbol{1}'_j \boldsymbol{u}_j \\ \boldsymbol{X}'_j \boldsymbol{u}_j \\ \tau_{ITT,X} T_j n_j - \tau_{CACE-T}[T_j \boldsymbol{1}'_j \boldsymbol{r}_j] \end{pmatrix}. \tag{C.4}$$

The $\hat{\boldsymbol{\Gamma}}_j(\hat{\boldsymbol{\xi}})$ matrix in (C.2) also applies to this estimator with the following changes: (i) $\boldsymbol{B} = (-T_j n_j \quad T_j n_j \quad T_j n_j (\overline{\overline{\boldsymbol{X}}}^1_W - \overline{\overline{\boldsymbol{X}}}^0_W))$ and (ii) $C = \sum_{i=1}^{n_j} T_j r_{ij}$ is now a scalar.

*C.3. Variance estimator if the weights are known*

If we assume the weights are known (a common approach used in practice), one can use a simple consistent variance estimator developed in Schochet et al. [40] for WLS (ratio) estimators for clustered designs that is asymptotically equivalent to the GEE estimator but performs better in simulations when the number of clusters is small. This estimator applies to all our considered estimators as it allows for general weighting schemes. The variance estimator, which is based on WLS residuals averaged to the cluster level, is as follows:

$$V\hat{a}r(\hat{\tau}) = \left(\frac{s^2(1)}{m^1} + \frac{s^2(0)}{m^0}\right), \text{where} \quad (C.5)$$

$$s^2(1) = \frac{1}{(m^1 - k\hat{p}_W - 1)(\widehat{\bar{w}}^1)^2} \sum_{j:T_j=1}^{m^1} \widehat{w}_j^2 \, (\bar{y}_{Wj} - \bar{\hat{y}}_{Wj})^2;$$

$$s^2(0) = \frac{1}{(m^0 - k(1-\hat{p}_W) - 1)(\widehat{\bar{w}}^0)^2} \sum_{j:T_j=0}^{m^0} \widehat{w}_j^2 \, (\bar{y}_{Wj} - \bar{\hat{y}}_{Wj})^2;$$

$\widehat{w}_j = \sum_{i=1}^{n_j} \widehat{w}_{ij}$ is the cluster weight; $\widehat{\bar{w}}^1 = \frac{1}{m^1}\sum_{j=1}^{m} T_j \widehat{w}_j$ and $\widehat{\bar{w}}^0 = \frac{1}{m^0}\sum_{j=1}^{m}(1-T_j)\widehat{w}_j$ are mean weights; $\hat{p}_W = \frac{1}{\sum_{j=1}^m \widehat{w}_j}\sum_{j=1}^{m} T_j \widehat{w}_j$ is the weighted proportion of treatments in the sample, and $\bar{y}_{Wj} = \frac{1}{\widehat{w}_j}\sum_{i=1}^{n_j}\widehat{w}_{ij}y_{ij}$ and $\bar{\hat{y}}_{Wj} = \frac{1}{\widehat{w}_j}\sum_{i=1}^{n_j}\widehat{w}_{ij}\hat{y}_{ij}$ are weighted cluster-level mean outcomes and predicted values. The free RCT-YES software (www.rct-yes.com) performs these calculations.

Note that without covariates, the GEE empirical sandwich estimator is $[\frac{(m^1-1)}{m^1}s^2(1) + \frac{(m^0-1)}{m^0}s^2(0)]$, so is very similar to (C.5). In practice, the GEE variance is often multiplied by a small sample correction term, $g$, such as $g = \left(\frac{m}{m-1}\right)\left(\frac{n-1}{n-2}\right)$, a common value in statistical software packages such as Stata [46]. The GEE expression for the model with covariates is more complex but is also asymptotically equivalent to (C.5).

**Appendix D. Detailed Simulation Methods and Results**

The data generating process in (19) of the main text was used for the simulations to generate potential service receipt decisions. The same two normally distributed covariates, $X_{ij1}$ and $X_{ij2}$, were included in the treatment and control models. The mean-zero covariates were generated independently as described in the main text using $\sigma_X^2 = \sigma_{uX}^2 + \sigma_{\varepsilon X}^2 = 0.15$ (see below), where we set $\sigma_{uX}^2 = 0.045$ and $\sigma_{\varepsilon X}^2 = 0.105$ so that 30 percent of the total variance of the covariates was due to variation between clusters. We selected the treatment group intercept, $\alpha_0^1$, so that the expected treatment group service receipt rate, $\bar{R}(1) = (\pi_{11} + \pi_{10})$, was 50 or 70 percent, and we then selected the control group intercept, $\alpha_0^0$, so that its expected service receipt rate, $\bar{R}(0) = (\pi_{11} + \pi_{01})$ was 10 or 20 percentage points lower than the rate for the treatment group. Specifically, we set $\alpha_0^1 = \ln(\bar{R}(1)/(1-\bar{R}(1)))$ and $\alpha_0^0 = \ln(\bar{R}(0)/(1-\bar{R}(0)))$.

The $\alpha_1^t$ and $\alpha_2^t$ parameters for $t \in \{1,0\}$ were then selected so that a one standard deviation change in the covariate would change the probability of service receipt by 5 or 10 percentage points—that we refer to as moderate or strong covariates. To implement this, we used the first-order approximation, $\alpha_1^t \approx \sigma_X \frac{\partial e^t(X)}{\partial X_1}/(\bar{R}(t)(1-\bar{R}(t)))$, for the first covariate and similarly for the second covariate, setting $\frac{\partial e^t(X)}{\partial X_1}$ to 0.05 or 0.10 and plugging in values for $\sigma_X$ and $\bar{R}(t)$ as discussed above. Under this approach, $\alpha_1^t = \alpha_2^t$, but they differed for the treatment and control models due to differences in the $\bar{R}(1)$ and $\bar{R}(0)$ values. Further, we allowed the covariate strength to be the same or different across the treatment and control group models.

Next, using the simulated covariates and model parameters, we calculated the propensity scores, $e_{ij}^t(X)$. We then used these propensity scores to generate the potential service receipt variables, $R_{ij}(t)$, to achieve the target $\bar{R}(1)$ and $\bar{R}(0)$ sample averages. This was done in two steps by (1) drawing a *Uniform*(0,1) random variable, $U_{ij}^t$, and (2) setting $R_{ij}(t) = 1$ if $e_{ij}^t(X) \geq U_{ij}^t$ and $R_{ij}(t) = 0$ otherwise. The $R_{ij}(1)$ and $R_{ij}(0)$ values were then used to allocate the sample to one of the four principal strata.

The data generating process in (20) of the main text was used to generate outcomes for the simulations. We parameterized all error variances in terms of the intraclass correlation coefficient for the control group, $ICC_0 = \sigma_u^2/(\sigma_u^2 + \sigma_\epsilon^2)$, the explained variance of the covariates in the control group, $R^2$, and the variance of the outcome in the control group, $\sigma_{y_0}^2$, which were all assumed to be the same across principal strata (so we omit the $(rr')$ subscripts). In addition, we assumed that the two model covariates were *iid* normal with the same variance, $\sigma_{X_v}^2 = \sigma_X^2$ and the same parameter values, $\gamma_v = 1$ for $v \in \{1,0\}$. Thus, because $R^2 = 2\gamma_v^2 \sigma_X^2/\sigma_{y_0}^2$, we have that $\sigma_X^2 = R^2 \sigma_{y_0}^2/2$. Let $\sigma_{y_0^*}^2 = \sigma_{y_0}^2 - 2\sigma_X^2$ and assume that $\sigma_\theta^2 = f\sigma_u^2$ for $f \geq 0$. We then have that $\sigma_u^2 = \sigma_{y_0^*}^2 ICC_0$, $\sigma_\epsilon^2 = \sigma_{y_0^*}^2(1 - ICC_0)$, and $\sigma_\theta^2 = f\sigma_{y_0^*}^2 ICC_0$. Further, we set $\sigma_{y_0}^2 = 1$ so that the outcomes are in effect size units based on the standard deviation of the control group, which is common practice. To apply the formulas and match real-world scenarios, we assumed that $ICC_0 = 0.10$, $f = 0.10$, and $R^2 = 0.30$. These assumptions yield simulation values of $\sigma_X^2 = 0.15$, $\sigma_u^2 = 0.07$, $\sigma_\theta^2 = 0.007$, and $\sigma_\epsilon^2 = 0.63$.

We ran 1,000 simulations for each specification, where the number of clusters was set to $m = 20$ or $80$, and the number of individuals, $n_j$, was selected from a *Uniform*(40,80) distribution. For each simulation round, we randomly assigned clusters to either the treatment or control group using a treatment assignment probability $(p)$ of 0.6. Assorted simulation results are presented in Tables D.1 and D.2.

**Table D.1. Assorted simulation results for the considered IPW estimators**

| Estimator and specification | True treatment effect[a] | Mean estimated treatment effect[a] | True standard error[b] | Mean estimated standard error[c] | Confidence interval coverage |
|---|---|---|---|---|---|
| $m = 80, \bar{R}(1) = .50, \bar{R}(0) = .40$, **strong treatment and moderate control logit covariates** | | | | | |
| $\hat{\tau}_{CACE-T}$ | .260 | .258 | .070 | .067 / .067 | .938 |
| $\hat{\tau}^1_{CACE-TC}$ | .157 | .156 | .098 | .095 / .098 | .933 |
| $\hat{\tau}^2_{CACE-TC}$ | .157 | .156 | .069 | .066 / .066 | .934 |
| $\hat{\tau}_{11}$ | .200 | .199 | .074 | .072 / .072 | .937 |
| $m = 20, \bar{R}(1) = .50, \bar{R}(0) = .40$, **strong treatment and moderate control logit covariates** | | | | | |
| $\hat{\tau}^{WLS}_{CACE-T}$ | .260 | .269 | .137 | .125 / .125 | .934 |
| $\hat{\tau}^{WLS,1}_{CACE-TC}$ | .157 | .172 | .188 | .182 / .188 | .961 |
| $\hat{\tau}^{WLS,2}_{CACE-TC}$ | .157 | .167 | .133 | .122 / .122 | .942 |
| $\hat{\tau}^{WLS}_{11}$ | .200 | .208 | .145 | .136 / .135 | .947 |
| $m = 80, \bar{R}(1) = .70, \bar{R}(0) = .50$, **moderate treatment and moderate control logit covariates** | | | | | |
| $\hat{\tau}_{CACE-T}$ | .250 | .258 | .066 | .066 / .066 | .948 |
| $\hat{\tau}^1_{CACE-TC}$ | .188 | .198 | .077 | .078 / .077 | .948 |
| $\hat{\tau}^2_{CACE-TC}$ | .188 | .196 | .065 | .066 / .066 | .947 |
| $\hat{\tau}_{11}$ | .200 | .207 | .069 | .069 / .069 | .944 |
| $m = 20, \bar{R}(1) = .70, \bar{R}(0) = .50$, **moderate treatment and moderate control logit covariates** | | | | | |
| $\hat{\tau}^{WLS}_{CACE-T}$ | .251 | .250 | .136 | .123 / .123 | .932 |
| $\hat{\tau}^{WLS,1}_{CACE-TC}$ | .189 | .188 | .157 | .150 / .157 | .950 |
| $\hat{\tau}^{WLS,2}_{CACE-TC}$ | .189 | .188 | .134 | .122 / .122 | .944 |
| $\hat{\tau}^{WLS}_{11}$ | .201 | .200 | .140 | .131 / .131 | .950 |

Notes. See main text and Appendix D for simulation details. The calculations assume WLS models with two covariates, a mean of $n_j = 60$ individuals per cluster (randomly selected from a *Uniform*(40,80) distribution), and a treatment group assignment rate of $p = .60$. The figures are based on 1,000 simulations.

WLS = weighted least squares.

[a]Treatment effects are measured in effect size (standard deviation) units.
[b]True standard errors are measured as the standard deviation of the estimated treatment effects across simulations.
[c]The first standard error adjusts for estimation error in the weights, while the second standard error does not.

**Table D.2. Assorted simulation results for the considered IPW estimators**

| Estimator and specification | True treatment effect[a] | Mean estimated treatment effect[a] | True standard error[b] | Mean estimated standard error[c] | Confidence interval coverage |
|---|---|---|---|---|---|
| $m = 80, \bar{R}(1) =.50, \bar{R}(0) =.40$, moderate treatment and moderate control logit covariates | | | | | |
| $\hat{\tau}_{CACE-T}$ | .260 | .258 | .068 | .068 / .068 | .953 |
| $\hat{\tau}^1_{CACE-TC}$ | .157 | .154 | .094 | .095 / .094 | .958 |
| $\hat{\tau}^2_{CACE-TC}$ | .157 | .155 | .067 | .067 / .066 | .956 |
| $\hat{\tau}_{11}$ | .200 | .198 | .072 | .072 / .072 | .952 |
| $m = 20, \bar{R}(1) =.50, \bar{R}(0) =.40$, moderate treatment and moderate control logit covariates | | | | | |
| $\hat{\tau}^{WLS}_{CACE-T}$ | .260 | .254 | .140 | .124 / .124 | .935 |
| $\hat{\tau}^{WLS,1}_{CACE-TC}$ | .157 | .149 | .194 | .181 / .193 | .952 |
| $\hat{\tau}^{WLS,2}_{CACE-TC}$ | .157 | .152 | .138 | .122 / .122 | .940 |
| $\hat{\tau}^{WLS}_{11}$ | .200 | .197 | .149 | .135 / .135 | .943 |
| $m = 80, \bar{R}(1) =.70, \bar{R}(0) =.60$, strong treatment and strong control logit covariates | | | | | |
| $\hat{\tau}_{CACE-T}$ | .249 | .249 | .068 | .066 / .066 | .947 |
| $\hat{\tau}^1_{CACE-TC}$ | .188 | .187 | .080 | .078 / .080 | .947 |
| $\hat{\tau}^2_{CACE-TC}$ | .188 | .187 | .068 | .066 / .066 | .950 |
| $\hat{\tau}_{11}$ | .199 | .199 | .071 | .069 / .069 | .948 |
| $m = 20, \bar{R}(1) =.70, \bar{R}(0) =.60$, strong treatment and strong control logit covariates | | | | | |
| $\hat{\tau}^{WLS}_{CACE-T}$ | .251 | .250 | .129 | .122 / .122 | .946 |
| $\hat{\tau}^{WLS,1}_{CACE-TC}$ | .189 | .190 | .151 | .149 / .150 | .958 |
| $\hat{\tau}^{WLS,2}_{CACE-TC}$ | .188 | .189 | .128 | .121 / .121 | .952 |
| $\hat{\tau}^{WLS}_{11}$ | .201 | .202 | .136 | .130 / .130 | .953 |

Notes. See main text and Appendix D for simulation details. The calculations assume WLS models with two covariates, a mean of $n_j = 60$ individuals per cluster (randomly selected from a *Uniform*(40,80) distribution) and a treatment group assignment rate of $p = .60$. The figures are based on 1,000 simulations.

WLS = weighted least squares.

[a]Treatment effects are measured in effect size (standard deviation) units.
[b]True standard errors are measured as the standard deviation of the estimated treatment effects across simulations.
[c]The first standard error adjusts for estimation error in the weights, while the second standard error does not.

**Appendix E. Additional Results for the Empirical Analysis**

This appendix presents additional results on the empirical analysis, including (i) a comparison of the baseline characteristics of the service recipients and nonrecipients in the treatment and control groups (Table E.1), (ii) a comparison of the lasso-selected characteristics of actual and weighted service recipients for each research group (Table E.2), and (iii) a comparison of the propensity score frequency distributions of the actual and weighted service recipients using a variant of the Shaikh et al. method [53] discussed in Appendix B (Figure E.1).

**Table E.1. Characteristics of service recipients and nonrecipients**

| Baseline characteristic (percentages) | Treatment group | | | Control group | | |
|---|---|---|---|---|---|---|
| | Service recipients | Service nonrecipients | Difference[a] | Service recipients | Service nonrecipients | Difference[a] |
| <u>Family and primary caregiver characteristics</u> | | | | | | |
| **Age of mother at birth of focus child** | | | | | | |
| Younger than 20 | 36.3 | 43.1 | -6.9 (5.8) | 40.5 | 38.1 | 2.3 (3.6) |
| 20-25 | 35.2 | 29.7 | 5.5 (5.7) | 29.5 | 33.2 | -3.7 (3.4) |
| 25 or older | 28.5 | 27.1 | 1.4 (5.4) | 30.1 | 28.7 | 1.4 (3.4) |
| **Highest grade completed** | | | | | | |
| Less than 9th grade | 11.9 | 19.0 | -7.1 (3.9) | 6.5 | 12.2 | -5.6 (2.2) |
| Grade 9-11 | 33.1 | 33.9 | -0.8 (5.6) | 36.2 | 37.5 | -1.3 (3.6) |
| High school or GED | 28.6 | 29.0 | -0.5 (5.4) | 29.8 | 25.1 | 4.6 (3.3) |
| More than 12th grade | 26.5 | 18.1 | 8.4 (5.3) | 27.5 | 25.1 | 2.3 (3.2) |
| **Race/Ethnicity** | | | | | | |
| White, non-Hispanic | 40.9 | 20.5 | 20.4 (5.9) | 49.9 | 37.2 | 12.7 (3.7) |
| Black, non-Hispanic | 31.3 | 43.7 | -12.4 (5.7) | 31.6 | 29.0 | 2.5 (3.4) |
| Hispanic | 24.1 | 28.9 | -4.8 (5.2) | 14.3 | 27.7 | -13.4 (3.1) |
| Other | 3.7 | 6.9 | -3.2 (2.3) | 4.3 | 6.1 | -1.8 (1.7) |
| **Primary occupation** | | | | | | |
| Employed | 25.5 | 21.4 | 4.1 (5.2) | 27.7 | 24.8 | 2.9 (3.2) |
| In school or training | 23.1 | 22.2 | 0.9 (5.1) | 23.2 | 20.5 | 2.7 (3.0) |
| Neither | 51.4 | 56.4 | -5.0 (6.0) | 49.1 | 54.7 | -5.6 (3.7) |
| **English language ability** | | | | | | |
| Primary language is English | 79.7 | 76.3 | 3.4 (4.9) | 85.2 | 76.0 | 9.2 (3.0) |
| Not English; speaks English well | 9.0 | 12.8 | -3.8 (3.5) | 7.9 | 11.0 | -3.1 (2.2) |
| Not English; does not speak English well | 11.3 | 10.9 | 0.4 (3.8) | 6.9 | 13.0 | -6.1 (2.3) |
| **Living arrangements** | | | | | | |
| Spouse | 27.2 | 21.6 | 5.6 (5.4) | 25.0 | 29.4 | -4.4 (3.3) |
| Other adults | 38.2 | 35.1 | 3.1 (5.9) | 36.1 | 42.3 | -6.2 (3.7) |
| No other adults | 34.6 | 43.2 | -8.6 (5.8) | 38.9 | 28.3 | 10.6 (3.5) |
| **Adult male present in household** | 41.5 | 32.4 | 9.1 (6.0) | 37.2 | 42.4 | -5.3 (3.7) |
| **Nonfocal children present in household** | | | | | | |
| Younger than 5 | 35.9 | 28.4 | 7.5 (5.8) | 35.8 | 37.7 | -1.9 (3.6) |
| 6-17 years old | 37.2 | 28.4 | 8.8 (5.9) | 35.4 | 36.0 | -0.6 (3.6) |
| **Number of moves in the past year** | | | | | | |
| 0 | 50.0 | 62.3 | -12.3 (6.0) | 43.6 | 48.7 | -5.1 (3.7) |
| 1 | 28.8 | 23.5 | 5.3 (5.4) | 23.5 | 28.0 | -4.5 (3.2) |
| 2+ | 19.0 | 10.1 | 8.9 (4.6) | 19.7 | 18.0 | 1.8 (2.9) |
| **Owns home** | 13.7 | 8.4 | 5.3 (4.1) | 10.4 | 11.8 | -1.4 (2.3) |

|  | Treatment group | | | Control group | | |
|---|---|---|---|---|---|---|
| Baseline characteristic (percentages) | Service recipients | Service nonrecipients | Difference[a] | Service recipients | Service nonrecipients | Difference[a] |
| **Household income as a percentage of poverty level** | | | | | | |
| Less than 33 | 25.0 | 23.0 | 2.0 (5.3) | 22.2 | 24.6 | -2.4 (3.2) |
| 33-67 | 27.7 | 20.3 | 7.4 (5.4) | 26.0 | 23.5 | 2.5 (3.2) |
| 67-99 | 20.9 | 14.9 | 6.0 (4.9) | 23.6 | 23.5 | 0.1 (3.2) |
| 100 or more | 11.9 | 14.9 | -2.9 (4.0) | 12.5 | 11.5 | 1.0 (2.4) |
| **Public assistance receipt** | | | | | | |
| TANF | 27.1 | 33.5 | -6.4 (5.4) | 29.5 | 25.9 | 3.7 (3.3) |
| SNAP | 46.7 | 45.2 | 1.5 (6.0) | 49.8 | 45.1 | 4.7 (3.7) |
| Medicaid | 76.3 | 70.0 | 6.3 (5.1) | 77.9 | 72.5 | 5.3 (3.2) |
| SSI | 6.8 | 8.4 | -1.6 (3.0) | 5.4 | 7.5 | -2.1 (1.8) |
| WIC | 89.2 | 86.9 | 2.4 (3.7) | 85.1 | 86.5 | -1.4 (2.5) |
| Public housing | 9.4 | 10.0 | -0.6 (3.5) | 11.0 | 7.5 | 3.5 (2.1) |
| Characteristics of the focal child | | | | | | |
| **Age (in months)** | | | | | | |
| Unborn | 25.5 | 9.5 | 16.1 (5.2) | 30.2 | 25.4 | 4.8 (3.3) |
| Less than 5 | 34.0 | 40.5 | -6.6 (5.8) | 31.3 | 33.3 | -2.1 (3.5) |
| 5 or more | 40.5 | 50.0 | -9.5 (6.0) | 38.5 | 41.2 | -2.7 (3.7) |
| **Male** | 49.9 | 47.3 | 2.6 (6.1) | 51.4 | 49.4 | 2.0 (3.7) |
| **First born** | 60.9 | 65.2 | -4.3 (5.9) | 60.8 | 60.0 | 0.8 (3.7) |
| **Birthweight less than 2,500 grams**[a] | 6.3 | 8.7 | -2.4 (2.9) | 4.6 | 5.0 | -0.4 (1.6) |
| **Born more than 3 weeks early** | 10.6 | 12.3 | -1.6 (3.8) | 8.8 | 8.1 | 0.7 (2.1) |
| **Stayed in hospital after birth** | 11.8 | 16.8 | -5.1 (3.9) | 12.2 | 9.6 | 2.5 (2.3) |
| **Child's health and development** | | | | | | |
| People concerned | 8.7 | 7.0 | 1.8 (3.4) | 13.8 | 9.2 | 4.6 (2.3) |
| Received evaluation | 3.5 | 4.1 | -0.5 (2.2) | 7.2 | 3.1 | 4.1 (1.5) |
| **Risk categories** | | | | | | |
| Established risks | 7.2 | 8.1 | -0.9 (3.2) | 7.6 | 5.8 | 1.9 (1.8) |
| Biological or medical risks | 11.4 | 12.2 | -0.8 (3.9) | 15.3 | 10.0 | 5.2 (2.4) |
| Environmental risks | 21.8 | 28.4 | -6.6 (5.1) | 24.7 | 24.4 | 0.3 (3.2) |

Notes. EHS intake forms. Service recipients are those who received home-based or child care services from any source. GED = General Educational Development; SNAP = Supplemental Nutrition Assistance Program; SSI = Social Security Income; TANF = Temporary Assistance for Needy Families; WIC = Special Supplemental Nutrition Program for Women, Infants, and Children.
[a] Standard errors are in parentheses.

**Table E.2. Lasso-selected characteristics of actual and weighted service recipients**

| | Recipients in the treatment condition | | | Recipients in the control condition | | |
|---|---|---|---|---|---|---|
| Baseline characteristic (percentages) | Actual treatments | Weighted controls[a] | Standardized difference[b] | Actual controls | Weighted treatments | Standardized difference[b] |
| Family and primary caregiver characteristics | | | | | | |
| **Highest grade completed** | | | | | | |
| Less than 9th grade | 11.9 | 9.5 | .073 | 6.5 | 8.3 | -.071 |
| Grade 9-11 | 33.1 | 36.9 | -.081 | 36.2 | 32.1 | .085 |
| High school or GED | 28.6 | 26.6 | .043 | 29.8 | 32.4 | -.057 |
| More than 12th grade | 26.5 | 27.0 | -.011 | 27.5 | 27.2 | .006 |
| **Race/Ethnicity** | | | | | | |
| White, non-Hispanic^ | 40.9 | 44.2 | -.067 | 49.9 | 47.1 | .056 |
| Black, non-Hispanic | 31.3 | 28.7 | .055 | 31.6 | 33.8 | -.048 |
| Hispanic | 24.1 | 21.8 | .054 | 14.3 | 16.3 | -.058 |
| **Living arrangements** | | | | | | |
| Spouse | 27.2 | 28.4 | -.027 | 25.0 | 23.3 | .039 |
| Other adults | 38.2 | 39.9 | -.034 | 36.1 | 34.3 | .037 |
| No other adults | 34.6 | 31.7 | .061 | 38.9 | 42.4 | -.072 |
| **Adult male present in household** | 41.5 | 41.4 | .003 | NA | NA | NA |
| **Nonfocal children present in household** | | | | | | |
| Younger than 5* | 35.9 | 38.0 | -.044 | NA | NA | NA |
| 6-17 years old | 37.2 | 36.6 | .012 | NA | NA | NA |
| **Number of moves in the past year** | | | | | | |
| 0 | 50.0 | 45.4 | .092 | NA | NA | NA |
| 1 | 28.8 | 26.9 | .043 | NA | NA | NA |
| 2+ | 19.0 | 19.1 | -.001 | NA | NA | NA |
| **Household income as a percentage of poverty level** | | | | | | |
| Less than 33 | 25.0 | 23.8 | .028 | 22.2 | 22.9 | -.017 |
| 33-67 | 27.7 | 25.4 | .052 | 26.0 | 28.3 | -.050 |
| 67-99 | 20.9 | 23.3 | -.059 | 23.6 | 20.2 | .079 |
| 100 or more | 11.9 | 11.9 | .001 | 12.5 | 12.3 | .005 |
| **Public assistance receipt** | | | | | | |
| TANF | 27.1 | 26.6 | .010 | 29.5 | 29.8 | -.005 |
| SNAP | 46.7 | 47.1 | -.007 | 49.8 | 49.5 | .005 |
| Medicaid^ | NA | NA | NA | 77.9 | 79.0 | -.027 |
| Characteristics of the focal child | | | | | | |
| **Age (in months)** | | | | | | |
| Unborn | 25.5 | 28.9 | -.078 | NA | NA | NA |
| Less than 5 | 34.0 | 31.8 | .046 | NA | NA | NA |
| 5 or more | 40.5 | 39.3 | .025 | NA | NA | NA |
| **Male*** | 49.9 | 50.3 | -.009 | 51.4 | 50.0 | .028 |

|  | Recipients in the treatment condition | | | Recipients in the control condition | | |
|---|---|---|---|---|---|---|
| Baseline characteristic (percentages) | Actual treatments | Weighted controls[a] | Standardized difference[b] | Actual controls | Weighted treatments | Standardized difference[b] |
| **Born more than 3 weeks early** | 10.6 | 8.2 | .080 | 8.8 | 10.9 | -.073 |
| **Stayed in hospital after birth** | 11.8 | 10.0 | .056 | 12.2 | 14.3 | -.067 |
| Got evaluation on health/development | NA | NA | NA | 7.2 | 5.9 | .049 |
| **Risk categories** | | | | | | |
| Biological or medical risks | 11.4 | 12.0 | -.019 | 15.3 | 14.8 | .013 |
| Environmental risks | 21.8 | 23.8 | -.049 | 24.7 | 22.5 | .050 |

Notes. Data come from the Early Head Start Evaluation intake forms. Service recipients are those who received home-based or child care services from any source. The figures pertain to all variables in the categories selected by lasso (even if lasso selected only a subset of the variables). GED = General Educational Development; SNAP = Supplemental Nutrition Assistance Program; TANF = Temporary Assistance for Needy Families; NA = not applicable because variable was not selected by lasso for the logit model.

[a]Weights are obtained from the fitted logistic regression model estimated using the other research group.

[b]Standardized differences are the differences in the covariates between the actual and weighted research groups divided by the standard deviation of the covariate for the actual group. None of the differences are statistically significant at the 5 percent level.

\* Interaction term included in the treatment group logit model
^ Interaction term included in the control group logit model

**Figure E.1. Propensity score frequency distributions using a variant of the Shaikh et al. approach**

a. Treatment group model

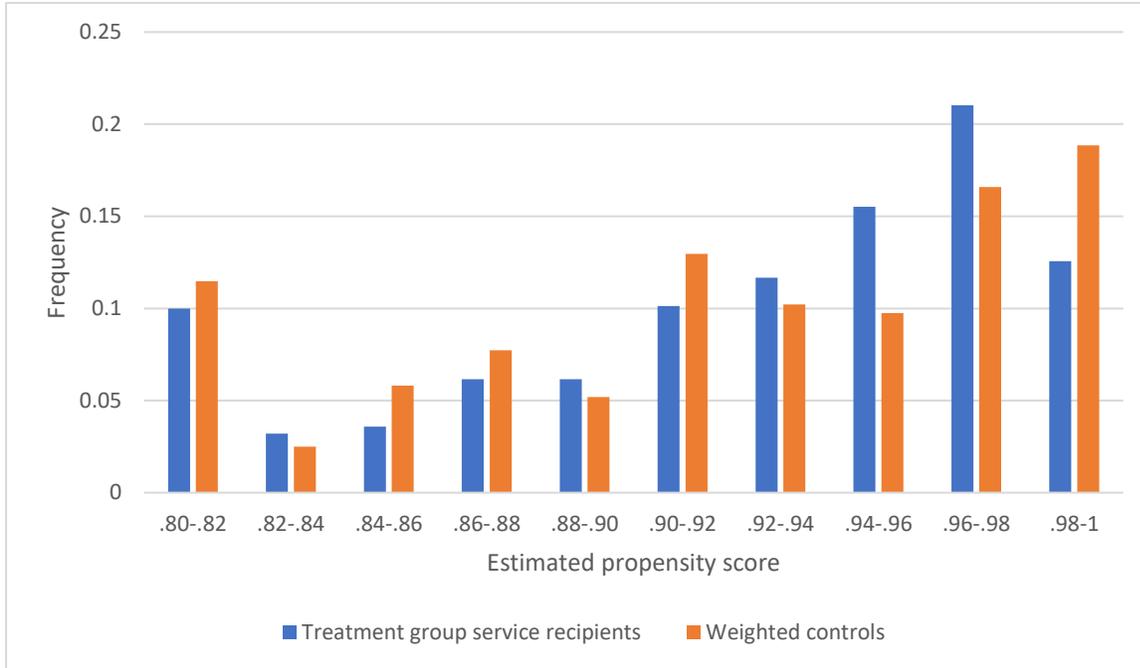

b. Control group model

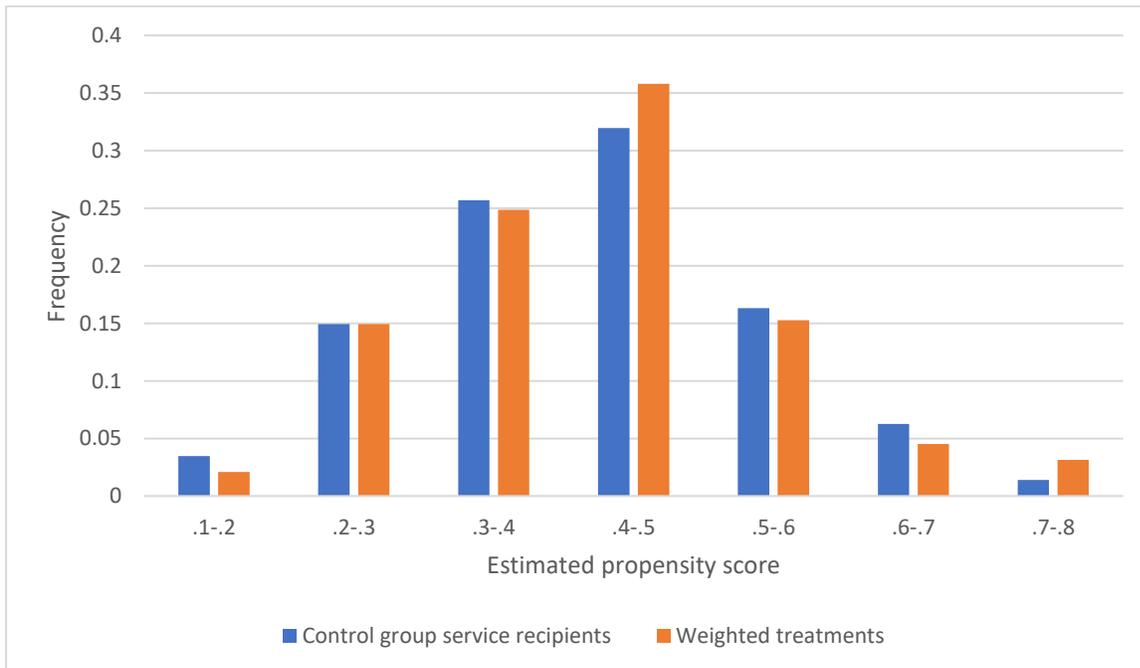